\documentclass{aa}

\usepackage{amssymb}
\usepackage{amsmath}
\usepackage{txfonts}
\usepackage{graphicx}
\usepackage{colortbl}
\usepackage[usenames]{xcolor}
\usepackage[bookmarks, colorlinks, breaklinks]{hyperref}
\hypersetup{linkcolor=blue,citecolor=blue,filecolor=black,urlcolor=blue} 
\usepackage{natbib}
\bibpunct{(}{)}{;}{a}{}{,} 
\newcommand{\bfx}{{\boldsymbol{x}}}

\begin{document}

\title{Hierarchical octree and $k$-d tree grids\\for 3D radiative transfer simulations}

\titlerunning{Octree and $k$-d trees in 3D radiative transfer}

\author{
W. Saftly
\and
M. Baes
\and
P. Camps   
}

\authorrunning{W. Saftly et al.}

\institute{ Sterrenkundig Observatorium, Universiteit Gent, Krijgslaan
  281, B-9000 Gent, Belgium\label{UGent} 
}

\date{\today}

\abstract{A crucial ingredient for numerically solving the
  three-dimensional radiative transfer problem is the choice of the
  grid that discretizes the transfer medium.  Many modern radiative
  transfer codes, whether using Monte Carlo or ray tracing techniques,
  are equipped with hierarchical octree-based grids to accommodate a
  wide dynamic range in densities.}%
{We critically investigate two different aspects of octree grids in
  the framework of Monte Carlo dust radiative transfer.  Inspired by
  their common use in computer graphics applications, we test
  hierarchical $k$-d tree grids as an alternative for octree grids. On
  the other hand, we investigate which node subdivision-stopping
  criteria are optimal for constructing of hierarchical grids.}%
{We implemented a $k$-d tree grid in the 3D radiative transfer code
  SKIRT and compared it with the previously implemented octree
  grid. We also considered three different node subdivision-stopping
  criteria (based on mass, optical depth, and density gradient
  thresholds). Based on a small suite of test models, we compared the
  efficiency and accuracy of the different grids, according to various
  quality metrics.}%
{For a given set of requirements, the $k$-d tree grids only require
  half the number of cells of the corresponding octree. Moreover, for
  the same number of grid cells, the $k$-d tree is characterized by
  higher discretization accuracy. Concerning the subdivision stopping
  criteria, we find that an optical depth criterion is not a useful
  alternative to the more standard mass threshold, since the resulting
  grids show a poor accuracy. Both criteria can be combined; however,
  in the optimal combination, for which we provide a simple
  approximate recipe, this can lead to a 20\% reduction in the number
  of cells needed to reach a certain grid quality. An additional
  density gradient threshold criterion can be added that solves the
  problem of poorly resolving sharp edges and strong density
  gradients.}%
{We advocate the use of $k$-d trees and the proposed combination of
  criteria to set up hierarchical grids for 3D radiative
  transfer. These recipes are straightforward for implementing and
  should help to develop faster and more accurate 3D radiative
  transfer codes.}
  

\maketitle
\section{Introduction}

The 3D radiative transfer problem is one of the
toughest challenges in computational astrophysics
\citep{2013ARA&A..51...63S}. Fortunately, significant progress has
been made in the past decade, with a steadily increasing number of
codes capable of dealing with the radiative transfer problem in a
general 3D geometry. The vast majority of these codes are based on
Monte Carlo techniques \citep[e.g.][]{2001ApJ...551..269G,
2003CoPhC.150...99W, 2004MNRAS.350..565H, 2006A&A...459..797P,
2006MNRAS.372....2J, 2008A&A...490..461B, 2011ApJS..196...22B,
2011A&A...536A..79R, 2013ApJS..207...30W}
or ray tracing mechanisms \citep[e.g.][]{1997A&A...325..135X,
2006ApJ...645..920S}.
For recent overviews of dust radiative transfer codes, for example, including a
discussion of the advantages, weaknesses, and optimization techniques,
see \citet{2011BASI...39..101W} and \citet{2013ARA&A..51...63S}.

A crucial aspect of radiative transfer simulations is the choice of
the grid on which the transfer medium is partitioned. To limit memory
usage and simulation run time, while at the same time keeping a high
effective grid resolution where necessary, non-uniform grids are
becoming a standard ingredient of most 3D radiative transfer codes. In
such grids, the grid cell size varies locally depending on the
physical conditions of the medium. While unstructured grids such as
Voronoi tesselations have their particular advantages
\citep[e.g.][]{2010A&A...515A..79P, 2010A&A...523A..25B, Camps2013},
most radiative transfer codes use structured, hierarchical grids. In
particular, hierarchical octree grids, in which nodes are recursively
partitioned into eight subnodes, are a popular choice
\citep[e.g.][]{2001A&A...379..336K, 2003CoPhC.150...99W,
  2004MNRAS.350..565H, 2006MNRAS.372....2J, 2008A&A...490..461B,
  2006A&A...456....1N, 2011A&A...536A..79R}.

In \citet{2013A&A...554A..10S}, we presented a detailed
investigation of how hierarchical octrees are used in Monte Carlo dust
radiative transfer. We investigated different methods of setting
up an octree (using either regular or barycentric subdivision of the
nodes) and different methods of traversing the tree, and showed
that the combination of regular node subdivision and neighbor list
search is the most efficient method for calculating paths through an
octree. This work was, however, far from the final word on the use of
octrees in radiative transfer, since it also pinpointed a
number of unsolved questions and problems.

One issue that needs additional consideration is the stopping
criterion used for constructing the octree grid. In
\citet{2013A&A...554A..10S}, we used a simple criterion that continues
the subdivision of each node until it contains a dust mass lower
than some preset threshold value. While this criterion is intuitive,
simple, and straightforward to implement, it has a number of potential
drawbacks. It cannot make a distinction between large cells with low
density and tiny cells with high density, while this distinction
should be made if we want to concentrate the subdivision of the cells
in regions with high optical depth. In addition, such a criterion has
difficulty with strong density gradients and sharp edges, because it
stops the subdivision of nodes with large gradients or sharp
edges relatively quickly, whereas high spatial resolution is
usually required in these regions.

Taking one step back, one might wonder whether octrees are the most
suitable hierarchical space for partitioning structures for radiative
transfer. In other words, is the systematic subdivision into eight
subnodes the best option, despite the straightforward implementation?
The subdivision into eight subnodes every time can result in many
cells with a mass content that is substantially lower than the
threshold. All of these almost empty cells have to be determined and
traversed, which could imply substantial memory and runtime
overheads. This problem might be alleviated by using $k$-d trees
instead of octrees. Similar to octrees, $k$-d trees are hierarchical
structures that recursively partition a given volume, but they split
each node into two rather than eight subnodes. Nowadays they are the
most popular space partitioning structure in computer graphics
applications \citep[e.g.][]{MacDonald1990, Havran2000, Reshetov2005,
  Wald2006, Shevtsov2007, Zhou2008}. To the best of our knowledge,
$k$-d trees have not been used in the context of astrophysical
radiative transfer simulations.

The goal of this paper is to investigate these issues using the 3D
Monte Carlo dust radiative transfer code SKIRT \citep{2003MNRAS.343.1081B,
2011ApJS..196...22B} and a set of challenging 3D test models. 
While our tests are performed within the specific framework of the Monte Carlo
technique applied to continuum radiative transfer in dusty systems, the results and conclusions
should be generally applicable to most if not all radiative transfer simulation codes.
This paper is structured as follows. In Sect.~{\ref{KDAlgorithms.sec}} we
present the $k$-d tree algorithm, and we discuss its
implementation in the SKIRT code. In Sect.~{\ref{comparison.sec}} we
compare the efficiency of $k$-d tree and octree based grids in the
context of radiative transfer simulations. In
Sect.~{\ref{StoppingCriteria.sec}} we investigate different
subdivision stopping criteria for both octrees and $k$-d trees, in
order to optimize the construction of radiative transfer
grids. Sect.~{\ref{Conclusion.sec}} sums up and gives an outlook on
further improvements and remaining issues.

\section{Radiative transfer on $k$-d tree grids}
\label{KDAlgorithms.sec}

\subsection{Properties of $k$-d trees}

A $k$-d tree is a hierarchical space-partitioning data structure,
originally introduced by \citet{Bentley1975}. It is a special case of
the class of binary space partitioning (BSP) trees, which all
partition space by recursively splitting each node in just two
subnodes (or children) by means of a hyperplane (a simple plane in 3D
space). BSP trees were developed in the late 1970s in the 3D computer
graphics community to rapidly access spatial information about objects
in a scene for rendering purposes \citep{Fuchs1979, Fuchs1980}, and
are now routinely applied in computer-aided geometric design and 3D
video games, amongst others. The characteristic property of $k$-d
trees is that each node-splitting plane (or hyperplane in the general
multidimensional case) is aligned with the axes of the initial
cuboidal starting node (the root node). The difference between the
octree and a $k$-d tree is that the octree divides a node into eight
subnodes using three splitting planes, whereas the $k$-d tree only
uses one splitting plane. So each octree can in principle be
represented as a $k$-d tree in which one level of subdivision in the
former is equivalent to three levels of subdivision in the
latter. Conversely, not all $k$-d trees can be represented by octrees.

\subsection{Implementation in SKIRT}

We have implemented a $k$-d tree dust grid structure in the SKIRT
Monte Carlo radiative transfer code \citep{2003MNRAS.343.1081B,
2011ApJS..196...22B}. The construction algorithm for the $k$-d tree
is completely analogous to the technique we used for the construction
of the octree grid structure in SKIRT, as explained in
\citet{2013A&A...554A..10S}. The tree is represented as a list of
nodes that starts with the cuboidal root node. The construction of the
tree consists of a single loop over all nodes in the list. For every
node in the list, we test whether it is to be subdivided in two
sub-nodes by comparing the ratio $\delta$ between dust mass in the
node (calculated using a Monte Carlo integration) and the total dust
mass in the system to a preset threshold value
$\delta_{\text{max}}$. If $\delta>\delta_{\text{max}}$, the node is
subdivided and the two subnodes are appended to the list. The
construction phase ends when the last node of the list is reached, and
the actual dust grid consists of all leaf nodes in the list, i.e.\ all
nodes that are not further subdivided.

In the construction of an octree, the subdivision point
of each node can in principle be chosen freely, and
\citet{2013A&A...554A..10S} investigated two possible options for this
(regular and barycentric subdivision). In a $k$-d tree, there is more
freedom, as both the orientation and the location of the splitting
plane can be chosen.

Concerning the orientation of the splitting plane, the canonical
method of constructing $k$-d trees uses a cyclic approach, where one
cycles through the axes as one moves down the three. In our 3D case
that means that the root node would have a splitting plane
perpendicular to the $x$-axis, its children one perpendicular to the
$y$-axis, its grandchildren one perpendicular to the $z$-axis, its
great-grandchildren one perpendicular to the $x$-axis, and so
on. Another option would be to select the orientation of the splitting
plane perpendicular to the axis which has the strongest density
gradient in the cell.

Concerning the location of the splitting plane, there are also
different options possible. The most obvious option is to use the
geometric centre of the node, which results in two subnodes of the
equal volume. Another option could be to split the cell such that the
two subnodes have equal mass. This option is similar to the
barycentric subdivision in octree grids explored by
\citet{2013A&A...554A..10S}, but, contrary to octrees, it is possible
to achieve exact equal-mass splitting for $k$-d trees. A third option
is to choose the splitting location such that the two subnodes have
equal mean density. This method has shown to give good results in
computer graphics ray-shooting simulations \citep{Wald2006}.

Based on extensive testing of these different options and our similar
experience with the octree subdivision options, we have found that the
standard combination of cyclic orientation and equal-volume split is
the most suitable method for radiative transfer simulations. In the
remainder of this paper, these are the options we have always
considered.

\subsection{Grid traversal}

Another crucial ingredient in our implementation is the choice of the
$k$-d tree grid traversal algorithm. A typical simulation involves
the calculation of millions of random straight paths through the dust
grid, along which the optical depth needs to be integrated. As the
grid cells in $k$-d trees are simple cuboids aligned to the principle
axes, this integration comes down to determining the ordered list of
dust cells that the path intersects. The most important step in the
grid traversal process is to determine the next cell along the path
which contains the exit position from the previous cell.

\citet{2013A&A...554A..10S} investigated in great detail three
possible algorithms for this step in the context of octree grids. It
was found that a neighbor list search method is very fast and
efficient and beats the traditional top-down method
\citep{Glassner1984}. In the neighbor list search method, each leaf cell
in the tree contains six different lists (one for each wall) with
pointers to the neighboring cells. As soon as a path exits a cell
through a given wall, the next cell is found searching the list of
neighboring cells. These lists can easily be constructed during the
grid construction with almost no computational overhead, and by
sorting the neighbors according to decreasing area of overlap, the
algorithm is extremely efficient. For all the simulations discussed in
this paper, we have used the neighbor list search algorithm for the grid
traversal.

One caveat is that this method cannot be used to locate the first cell
along a path, i.e.\ find the cell that corresponds to a new emission
event. This operation needs to be done using top-down search. As a
$k$-d tree subdivides each node into two instead of eight sub-nodes, the
depth of a $k$-d tree is typically a factor three higher than the depth
of the corresponding octree. The number of operations to descend a
$k$-d tree is, however, roughly equal in both cases, as one test
only needs to be done per level in the $k$-d tree, compared to three tests
in the octree. Moreover, as the top-down search only needs to be performed
once per path, this does not affect the efficiency of the
algorithm.

\section{Comparison between octree and $k$-d tree grids in radiative transfer simulations}
\label{comparison.sec}

\subsection{Test models}

In order to compare the efficiency and accuracy of the $k$-d tree grid
structure in radiative transfer simulations to the octree version, we
have considered the same test models as in
\citet{2013A&A...554A..10S}. The first model is a three-armed
logarithmic spiral structure. This model is completely analytical and
is inspired by the spiral galaxy models used by
\citet{2000A&A...353..117M} and \citet{2012ApJ...746...70S}.  The
second model is a model for the central region of an active galactic
nucleus (AGN), and consists of a central, isotropic source surrounded
by a number of compact and optically thick clumps embedded in a smooth
interclump medium.  It is similar to the AGN torus models presented by
\citet{2012MNRAS.420.2756S}. Finally, the third test model is a
numerical spiral galaxy model created by means of an SPH
hydrodynamical simulation. The galaxy model we consider is the 1 Gyr
snapshot of model run number 6 from \citet{2012MNRAS.422.2609R}.

\subsection{Grid quality measures}

In order to test our grids and know which grid is better, we have to
define quality metrics on which we base our
decisions. Unfortunately it is not trivial, if not at all impossible,
to define unique characteristics of a "good" grid. In broad terms, a
good grid should satisfy a number of characteristics.

A first important metric is the total number of cells
$N_{\text{cells}}$. For a given dust mass threshold
$\delta_{\text{max}}$, the ideal grid has as few cells as possible:
the memory consumption scales directly with $N_{\text{cells}}$ and
also the simulation run time is expected to increase with an
increasing number of cells. In simulations where not the shooting of
photons through the grid, but the calculation of the temperature
distribution or the dust emissivity in each cell is computationally
the most expensive operation, the simulation run time scales directly
with $N_{\text{cells}}$. In many radiative transfer simulations, however,
e.g.\ in pure scattering problems, the grid traversal is
computationally the most expensive operation. This grid traversal is
roughly proportional to $\langle N_{\text{cross}}\rangle$, the average
number of cells crossed along a single path, so this is also an
important metric to take into account. The barycentric grids explored
by \citet{2013A&A...554A..10S} are a prime example demonstrating
that the two metrics $N_{\text{cells}}$ and $\langle
N_{\text{cross}}\rangle$ are not always directly related.

The combination of $N_{\text{cells}}$ and $\langle
N_{\text{cross}}\rangle$ doesn't tell the entire story: even if
two grids have the same number of cells and/or the same average number
of cells crossed along a path, one can still be considered better than
the other, because it has a cleverer placement of the cells and hence
a more accurate discretization of the dust density field. Therefore,
we have also considered two grid quality metrics that measure the
accuracy of the discretization.

First, we have evaluated the difference between the true density
$\rho_{\text{t}}(\bfx)$ and the grid density $\rho_{\text{g}}(\bfx)$
in a large number of positions $\bfx$, uniformly sampled from the dust
grid volume. Here, $\rho_{\text{t}}(\bfx)$ is the value of the density
from the input dust density field, and $\rho_{\text{g}}(\bfx)$ is the
value of the uniform density in the grid cell that contains the
position $\bfx$. Subsequently, we calculate the standard deviation,
$\Delta\rho$, of this distribution of
$\rho_{\text{t}}-\rho_{\text{g}}$ values and use it as a metric to
measure the accuracy of the discretization of the grid. Note that
there is no absolute value for this metric that can decide whether a
grid is a "good" grid: for each model we can have a different value
for this number, depending on the complexity of the geometry and the
total mass and distribution of the dust. But it is a useful metric to
compare two different grids corresponding to the same model: the grid
with the lower value of $\Delta\rho$ is more accurate than the other.

As a final quality metric we use a similar standard deviation, but now
corresponding to the optical depth. The rationale behind this metric
is that the dust density is not the ultimate quantity for which we
need the discretization. In the end, the goal of any radiative
transfer problem is to solve for the intensity of the radiation field,
and our grid should be optimized so that the refinements follow the
changes in this quantity \citep{2013ARA&A..51...63S}. But this is
extremely challenging, as we do not know the radiation field a
priori. One quantity that might follow the distribution of the
radiation field intensity more closely than the density is the optical
depth. We choose a large number of random straight paths through the
grid volume, with start and end points distributed uniformly across the dust grid volume.
For each path we evaluate the difference between the theoretical
optical depth $\tau_{\text{t}}$ (calculated by numerically integrating
the true density along the path) and the grid optical depth
$\tau_{\text{g}}$ (calculated on the grid). Subsequently, we calculate
the standard deviation $\Delta\tau$ for the distribution of
$\tau_{\text{t}}-\tau_{\text{g}}$. Again, the values of $\Delta\tau$
can differ widely between different models, but it is a useful metric
to compare the quality of two grids that correspond to the same model.

\subsection{Results}

\begin{figure*}
\centering
\includegraphics[width=0.49\textwidth]{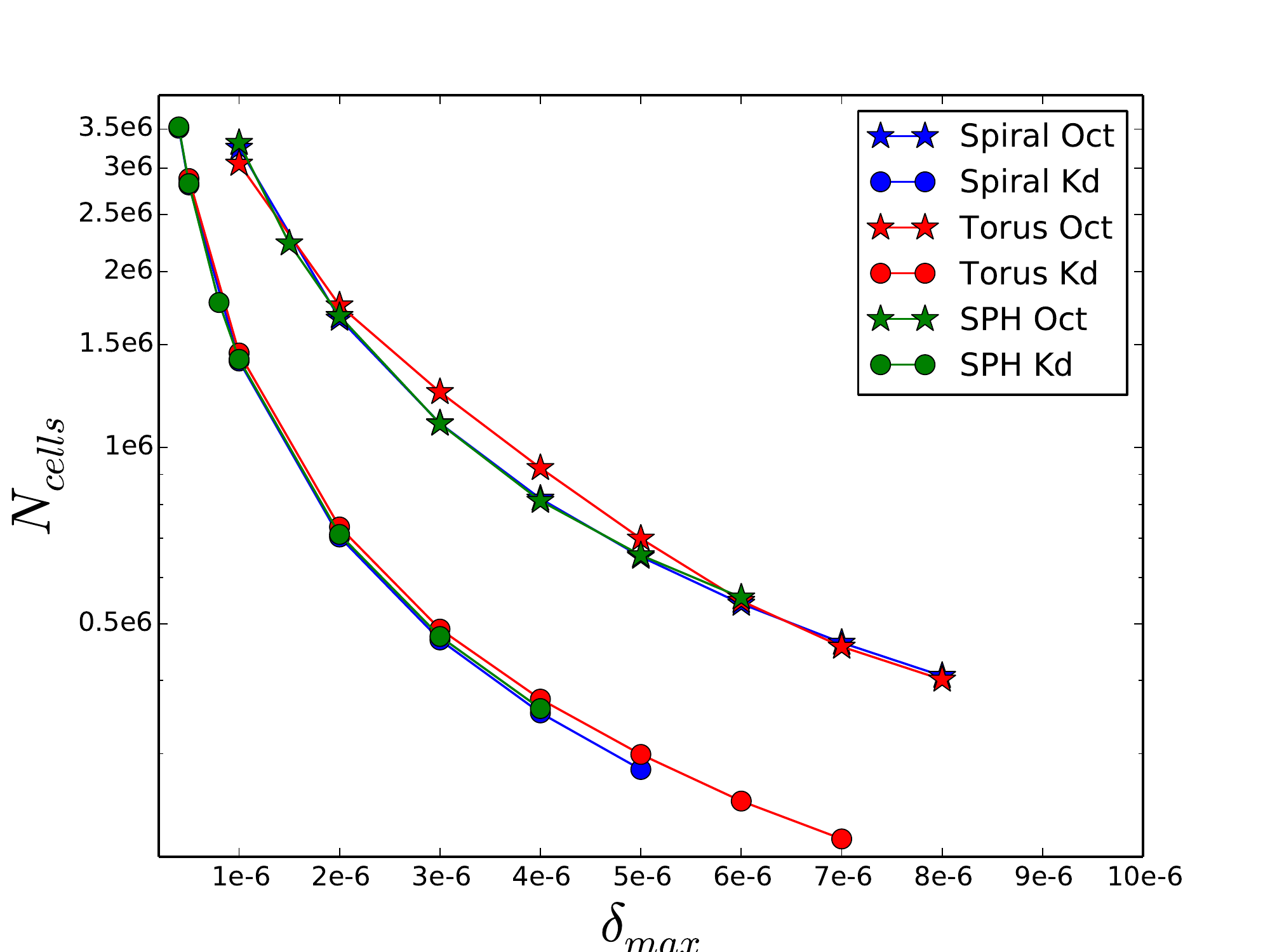}
\includegraphics[width=0.49\textwidth]{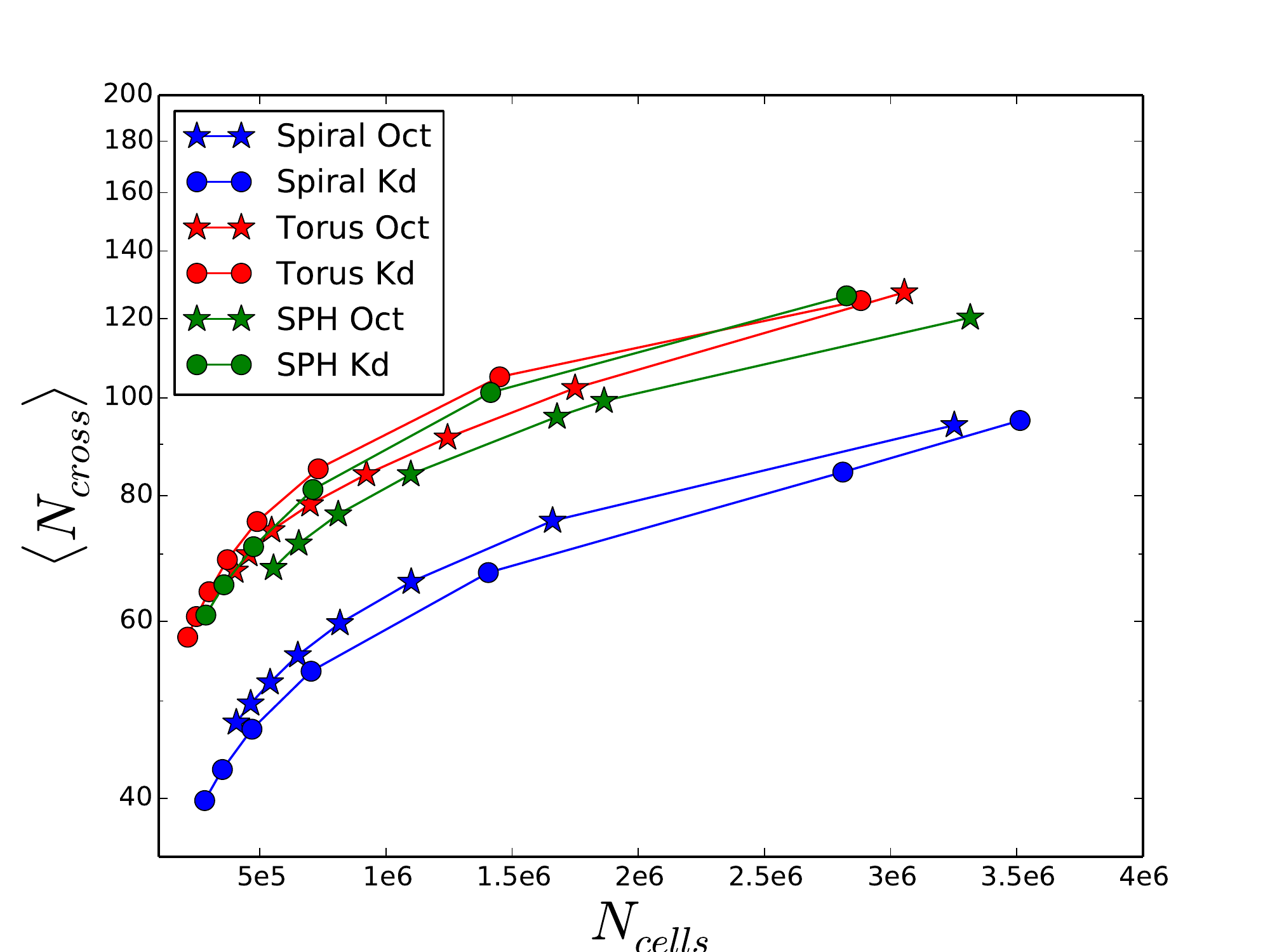}
\includegraphics[width=0.49\textwidth]{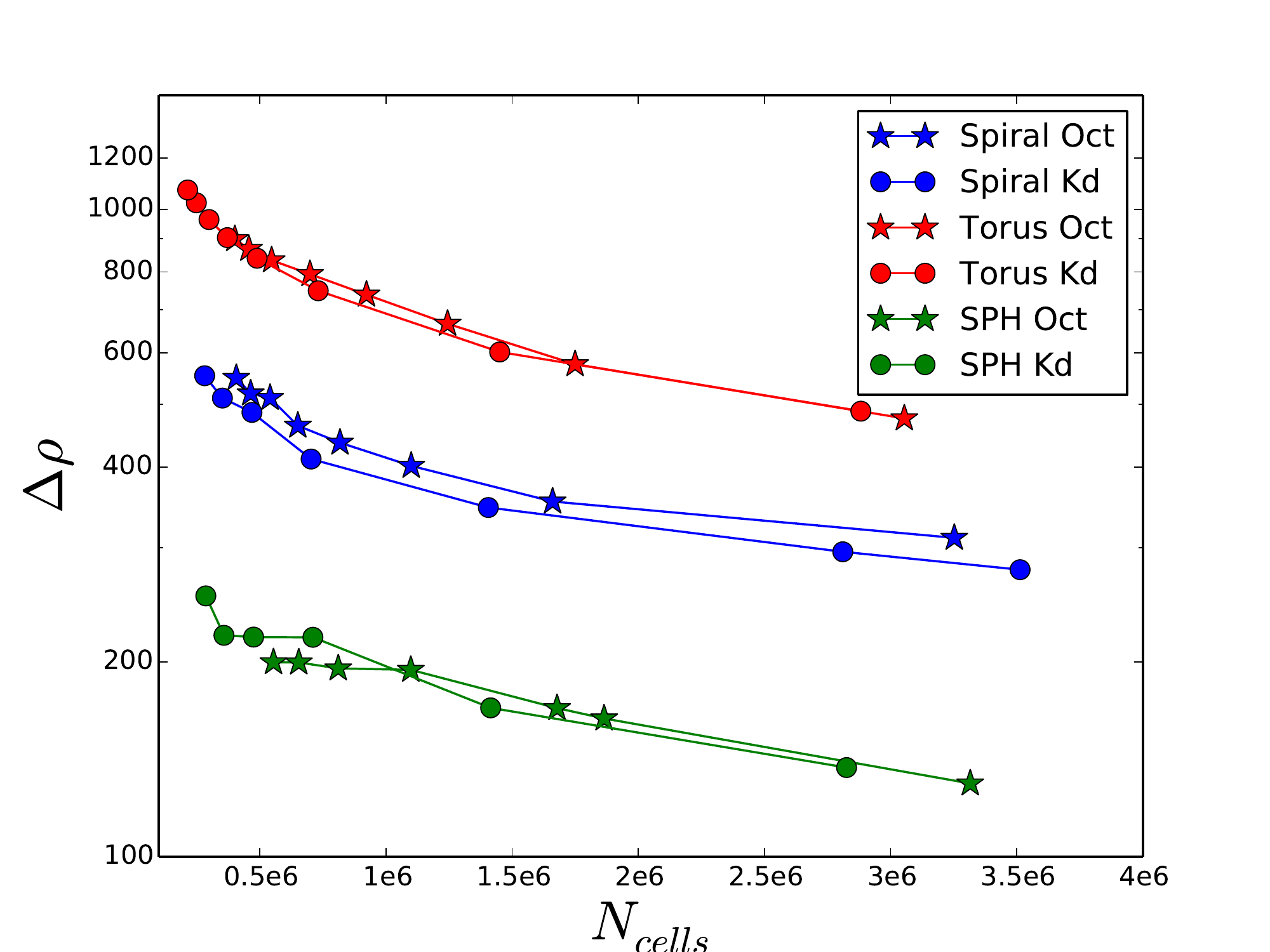}
\includegraphics[width=0.49\textwidth]{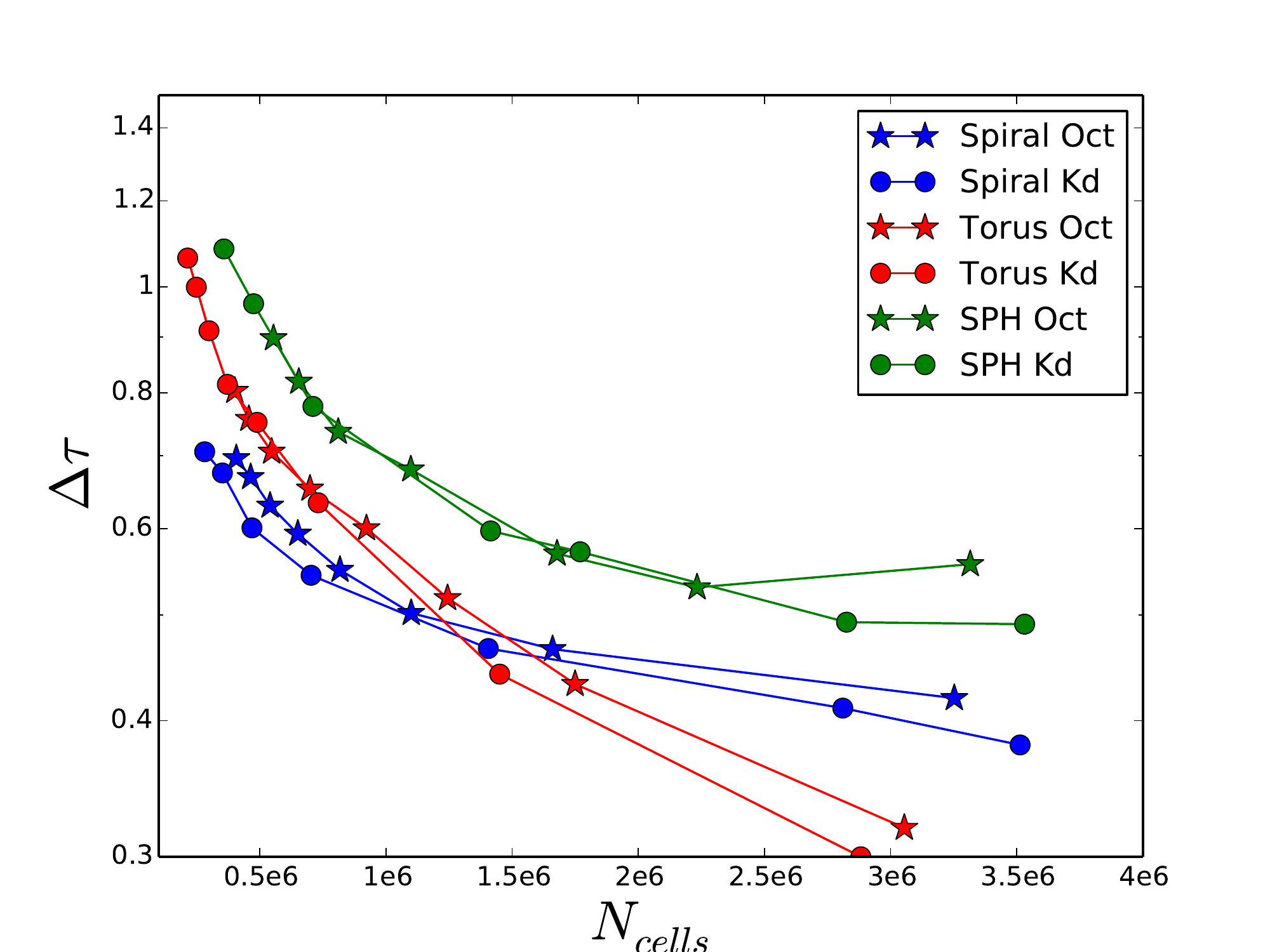}
\caption{Comparison of the grid quality metrics for octree and
  $k$-d-tree based grids of our three test cases (stars correspond to
  octree grids, dots to $k$-d tree grids, and the different colors
  correspond to the three different test models). The top left panel
  shows $N_{\text{cells}}$, the total number of cells in the grid, as
  a function of the threshold mass fraction $\delta_{\text{max}}$. For
  a given threshold value, octree grids require up to more than double
  the number of cells compared to the corresponding $k$-d tree. The
  other panels show the average number of cells crossed per path
  $\langle N_{\text{cross}}\rangle$, the density quality metric
  $\Delta\rho$, and the optical depth quality metric $\Delta\tau$ as a
  function of $N_{\text{cells}}$. }
\label{comparing.fig}
\end{figure*}

Fig.~{\ref{comparing.fig}} shows the results of a comparison of
octree- and $k$-d-tree based SKIRT simulations of our three test
cases. The top left panel shows $N_{\text{cells}}$, the total number
of cells in the grid, as a function of the threshold mass fraction
$\delta_{\text{max}}$. Impressively, for a fixed value of
$\delta_{\text{max}}$, the $k$-d tree generates only half as many
cells as the octree, irrespective of the chosen model. This alone
already strongly advocates for the use of $k$-d trees in radiative
transfer simulations.

In the top right panel of Fig.~{\ref{comparing.fig}} we show the
average number of cells crossed per path. If we would have plotted
this quantity as a function of the threshold mass fraction
$\delta_{\text{max}}$, the $k$-d tree would easily beat the octree
grid, as the former contains only half the number of cells of the
latter. If we plot $\langle N_{\text{cross}}\rangle$ as a function
of $N_{\text{cells}}$, the picture is more mixed: the $k$-d tree
performs slightly better than the octree for the logarithmic spiral
galaxy models, but slightly worse for the other two models. Averaging
out, we find that the average straight path through the $k$-d tree
crosses roughly the same number of cells as the average path through
the octree with the same total number of cells (with a more relaxed
value of $\delta_{\text{max}}$).

Looking at the quality of the grids, we find that, for a fixed
$\delta_{\text{max}}$, the octree grids have a more accurate
discretization than the corresponding $k$-d trees. This is no
surprise, as they contain up to double as many cells. On the other
hand, if we compare the quality of the octree and $k$-d tree grids for
 fixed values of $N_{\text{cells}}$ (bottom panels of
Fig.~{\ref{comparing.fig}}), the results turn around: for a fixed
number of cells, the $k$-d tree grids typically correspond to lower
values of $\Delta\rho$ and $\Delta\tau$ compared to the octree
grids. This means that, for a certain required density or optical
depth grid quality, we can generate a $k$-d tree with roughly 20\%
fewer cells compared to an octree grid. The general conclusion of this
investigation is that the $k$-d tree based grids outperform the octree
grids in all grid quality measures. As they are as simple to implement
as the more widely used octree grids, we recommend their use in
radiative transfer codes.

\section{Subdivision stopping criteria for hierarchical tree construction}
\label{StoppingCriteria.sec}

\subsection{Dust mass and optical depth criteria}

\begin{figure*}
\centering
\includegraphics[width=0.33\textwidth]{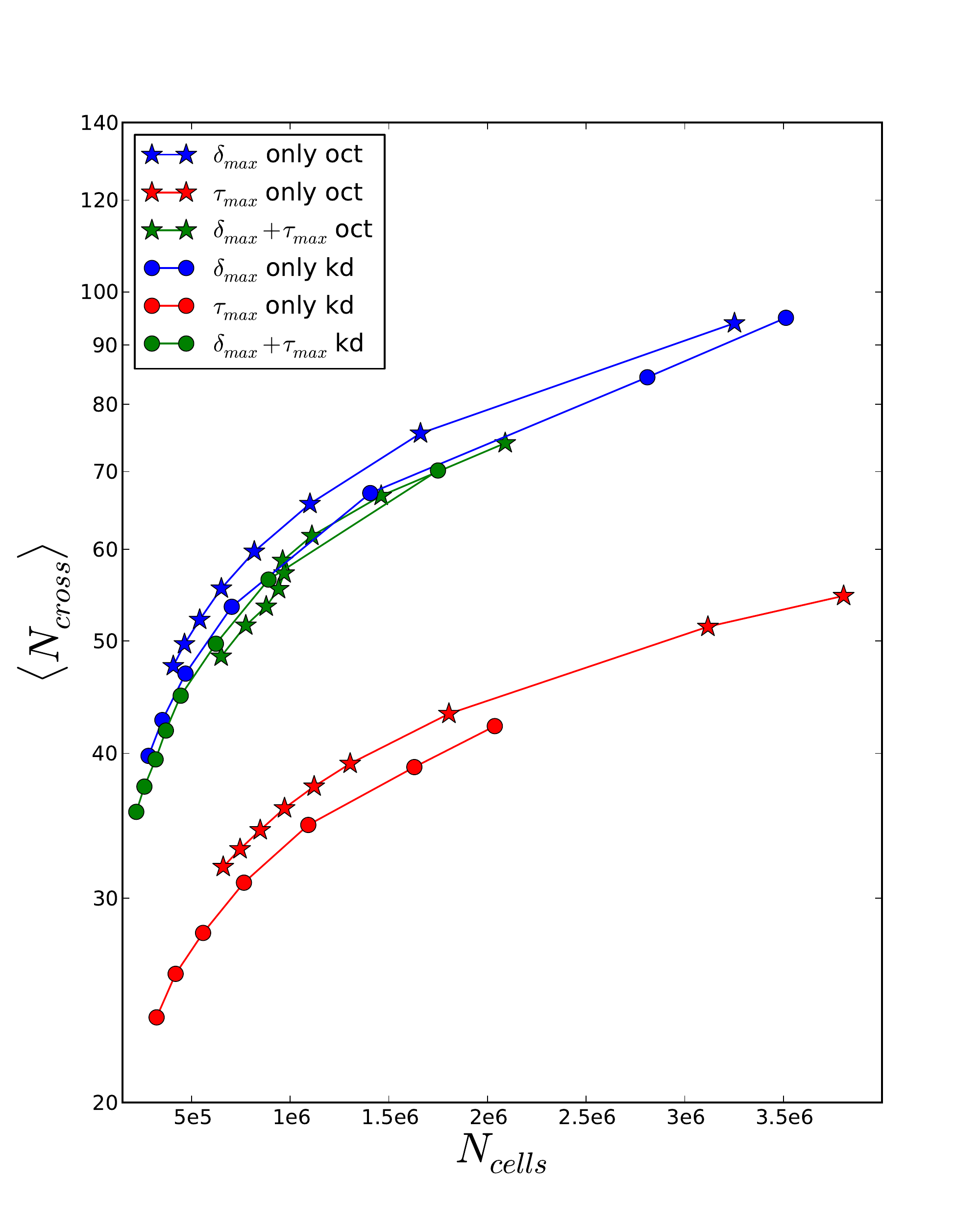}
\includegraphics[width=0.33\textwidth]{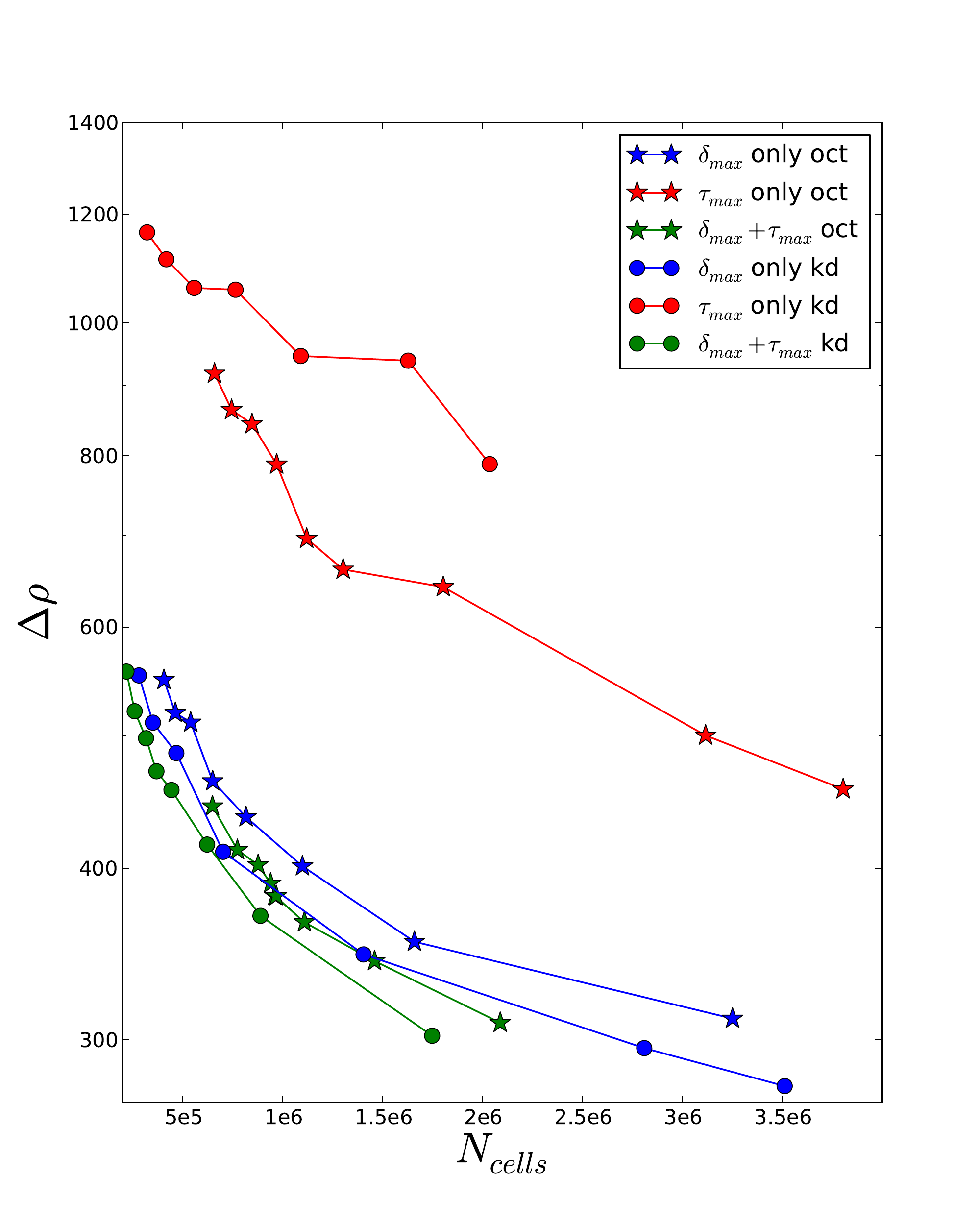}
\includegraphics[width=0.33\textwidth]{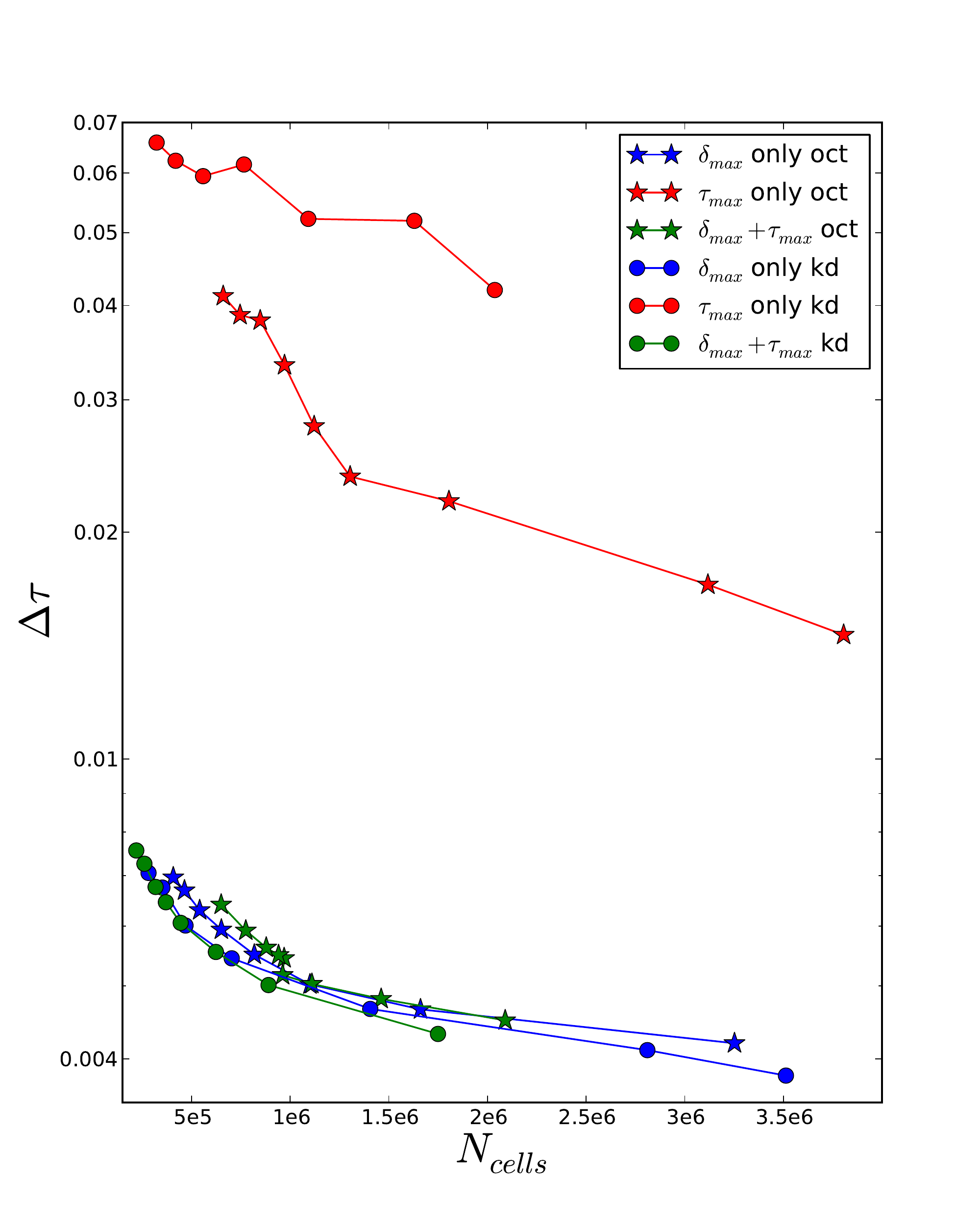}
\caption{Comparison of the grid quality metrics for octree (stars) and
  $k$-d-tree (dots) based grids of the spiral galaxy model, using
  different node subdivision stopping criteria. The blue lines
  correspond to a mass threshold only, the red lines lines to an
  optical depth threshold only, and the green lines to a combination
  of both criteria. The different panels represent the average number of cells crossed per path $\langle
  N_{\text{cross}}\rangle$, the density
  quality $\Delta\rho$, and the optical depth quality metric $\Delta\tau$
 as a function of the total number of cells
  in the grid, $N_{\text{cells}}$.}
  \label{Differences.fig}
\end{figure*}

Ideally, the spatial dust grid used in radiative transfer simulations
should be optimized to discretize the intensity of the radiation
field, the fundamental quantity that is the objective of radiative
transfer studies. Obviously, this quantity is not known at the
beginning of a radiative transfer simulation, which makes the task to
set up the grid hard: the challenge is to construct the best possible
grid without a priori knowledge of the radiation field. The most
obvious choice is to base the grid on the dust density field, but the
question still remains on which criterion we should base the
discretization.

For the construction of both the octree and the $k$-d tree based dust
grids we have used so far, we have adopted a simple mass-based
criterion to stop the recursive subdivision of the nodes. As long as
the fractional dust mass $\delta = M/M_{\text{tot}}$ in a node is
larger than a preset threshold value $\delta_{\text{max}}$, the node
is further subdivided. Here, the mass in a node is determined by
evaluating the true input density at $N_{\text{ran}}$ randomly
generated positions $\bfx_i$ within the node, i.e.
\begin{equation}
	M 
	=
	\frac{V}{N_{\text{ran}}}
	\sum_{i=1}^{N_{\text{ran}}} \rho_{\text{t}}(\bfx_i) 
\end{equation}
with $V$ the volume of the node. This criterion ensures that there
will be more subdivision, and hence higher resolution, in regions with
higher density, which is exactly what is desired. Moreover, the
criterion is very easy to implement. One weakness of this construction
algorithm is that does not distinguish between large cells with a low
dust density and small cells with a large density. This implies that
different cells, even if they contain the same dust mass, can have
strongly different volumes or densities. In order to resolve the
crucial regions where the radiation field changes most rapidly, it
probably makes sense to subdivide those cells with a high density
slightly more than the ones with a lower density, even if they contain
the same mass. This inspired us to look for another criterion that
can be used instead of, or in addition to, the mass criterion.

There are different possible options that can be explored to stimulate
the subdivision of nodes with a high density. All of these,
essentially, use a combination of the mass and the volume
of a node, and these quantities are easy to calculate during the
construction of the tree. We have opted for an optical depth
criterion, where we continue the subdivision if the optical depth of a
node is larger than a threshold $\tau_{\text{max}}$. For the optical
depth of a node, we use the maximum optical depth through the
node. For cuboidal nodes (as we have in our octree and $k$-d tree
grids), with sides $a$, $b$ and $c$, total mass $M$ and opacity
$\kappa$, the optical depth is
\begin{equation}
	\tau
	=
	\kappa\,M\,\frac{\sqrt{a^2+b^2+c^2}}{abc}
\end{equation}
We prefer this criterion above, for example, a criterion based on the
mean density, as the optical depth is a quantity that has a direct
link to the radiative problem (changes in the radiation field are
proportional to the optical depth in the optically thin cases).

A first question to consider is whether this optical depth criterion
can be used an alternative for the mass criterion we have used so
far. In Fig.~{\ref{Differences.fig}} we show the results of test
simulations based on the logarithmic three-armed spiral galaxy
model. For both the octree grid (stars) and the $k$-d tree grid
(dots), we constructed a set of dust grids using only a mass
criterion (blue lines) and using only an optical depth criterion
(red lines). The three different panels compare the grid metrics
$\langle N_{\text{cross}}\rangle$, $\Delta\rho$ and $\Delta\tau$ as a
function of the total number of grid cells $N_{\text{cells}}$. It is
clear from this figure that using only the optical depth criterion is
not a sensible option. For a fixed $N_{\text{cells}}$, an optical depth
based grid has a stronger subdivision in the high density regions
compared to a grid with mass based subdivision, but, unavoidably, this
also results in a larger number of large cells. The positive result is
that the former grids are faster, in the sense that the average path
crosses fewer cells. However, this advantage does not weigh up against
the loss of accuracy: the optical depth based grids have a much poorer
accuracy than the mass based grids, as measured by both $\Delta\rho$
and $\Delta\tau$. To achieve the same accuracy, around an order of
magnitude more cells are required.

\begin{figure*}
\centering
\includegraphics[width=0.49\textwidth]{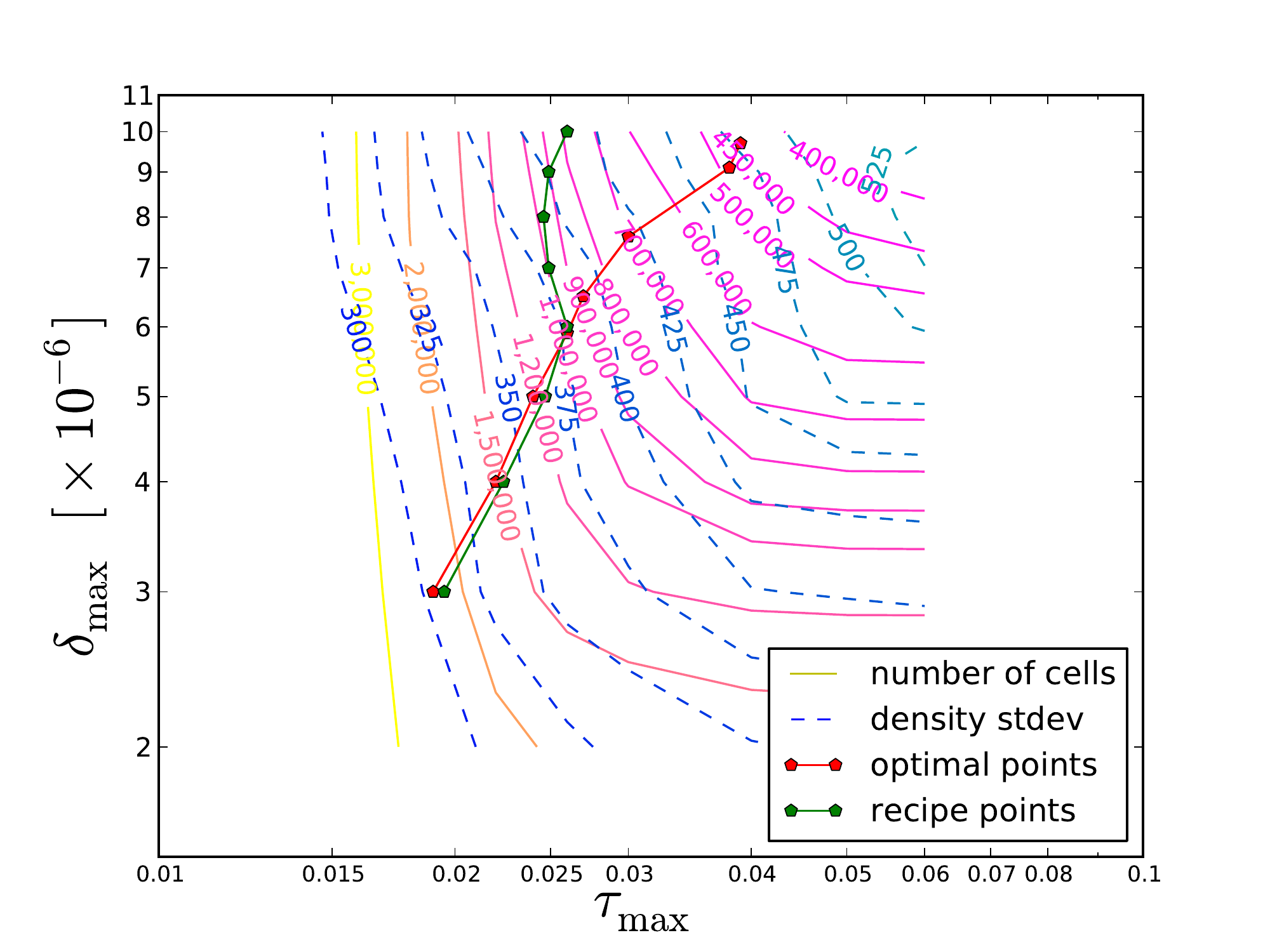}
\includegraphics[width=0.49\textwidth]{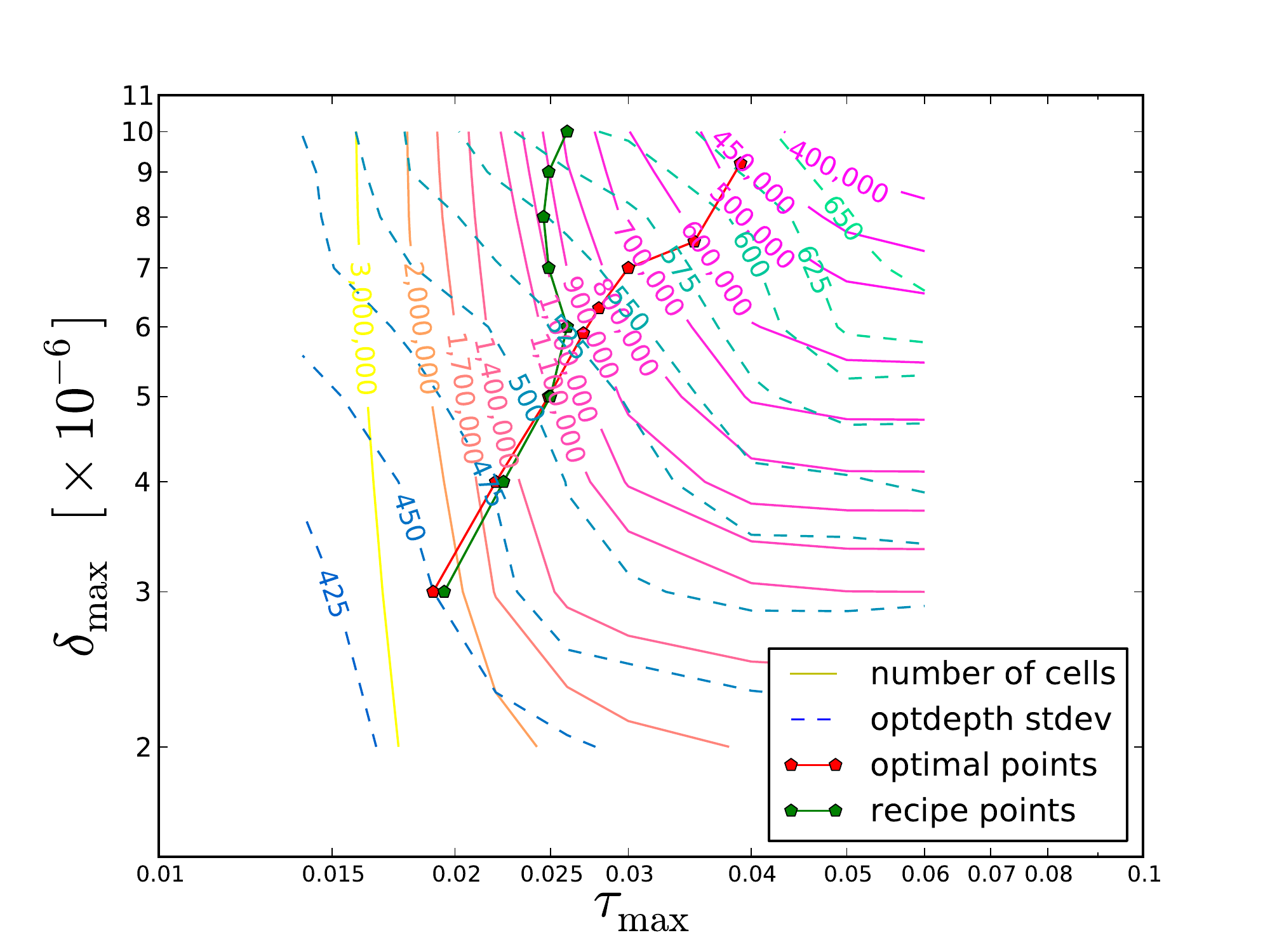}
\includegraphics[width=0.49\textwidth]{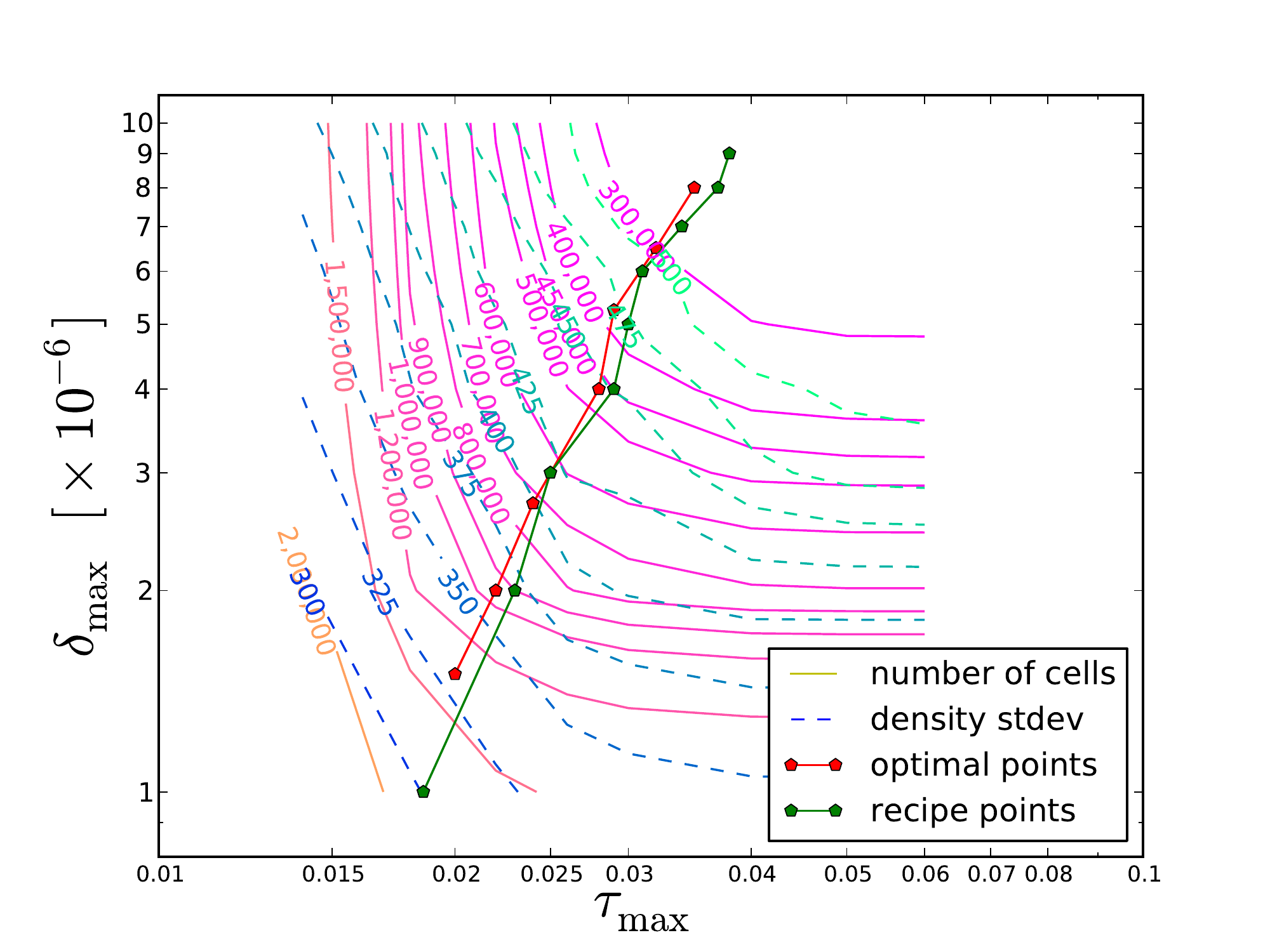}
\includegraphics[width=0.49\textwidth]{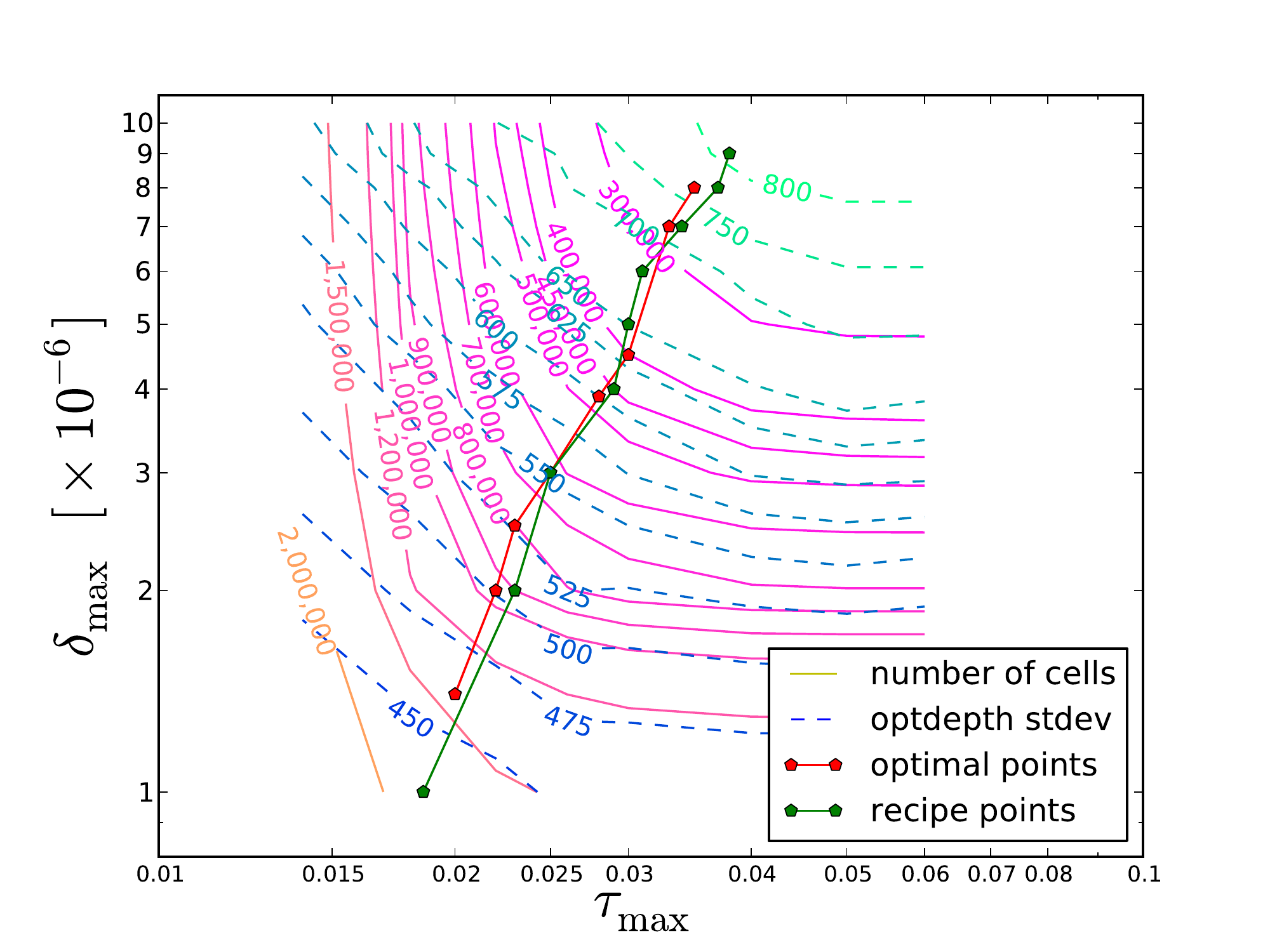}
\caption{Contour plots illustrating different properties of octree
  grids (left panels) and $k$-d tree grids (right panels), corresponding as a
  function of the $\delta_{\text{max}}$ and $\tau_{\text{max}}$
  threshold values. The solid lines in each panel correspond to lines
  of equal numbers of grid cells, whereas the dashed lines correspond
  to iso-quality contours, corresponding to the density quality metric
  $\Delta\rho$ (top panels) and the optical depth quality metric
  $\Delta\tau$ (bottom panel).}
\label{Contours.fig}
\end{figure*}

This does not imply that the optical depth criterion is useless, as it
can also be combined with the mass criterion. In
Fig.~{\ref{Differences.fig}} we also show the results of a grid that
uses a combination (green lines) of both criteria (details on how these two criteria
are combined are given below). Here we clearly see that the
combination is advantageous: by combining both criteria, we can
improve the accuracy of the grid, as measured by both $\Delta\rho$ and
$\Delta\tau$, and we decrease the average number of cells crossed on a
path, resulting in a speed-up of the simulation.

A useful tool to understand the complex interplay between these two
quantities are contour plots that display the total number of grid
cells and the different grid quality metrics as a function of the
stopping criteria $\tau_{\text{max}}$ and $\delta_{\text{max}}$.
Fig.~{\ref{Contours.fig}} shows such plots for the logarithmic
three-armed spiral galaxy model, for both the octree and $k$-d tree
grids. In each of these panels, the stopping criteria become more
stringent towards the bottom-left corner. Consequently, the number of
cells and the quality increase towards that same corner.

From these plots, we can identify the best combination of the stopping
criteria to construct a grid with a certain quality metric. Assume,
for example, that we want to construct an octree based grid with a
density quality metric $\Delta\rho=375$ (recall that the absolute
value is not relevant). We can then look at the top left plot, which
displays the contours of $N_{\text{cells}}$ and $\Delta\rho$ as a
function of $\delta_{\text{max}}$ and $\tau_{\text{max}}$. The contour
corresponding to $\Delta\rho=375$ connects all points in the
parameters space that correspond to grids with the same quality. We
can, for example, create such a grid by setting
$\delta_{\text{max}}=3\times10^{-6}$ and $\tau_{\text{max}} = 0.033$,
and looking at the $N_{\text{cells}}$ contours, we see that this grid
will contain about 1.2 million grid cells. We can relax the mass
criterion to $\delta_{\text{max}}=6\times10^{-6}$ and tighten the
optical depth criterion to $\tau_{\text{max}} = 0.026$: this will
yield a grid with the same density quality, but the number of grid
cells reduces to about 1 million. If we relax the mass threshold even
further to $\delta_{\text{max}}=10^{-5}$ and tighten the optical depth
threshold to $\tau_{\text{max}} = 0.022$, we obtain another grid with
the same density quality, but the number of cells increases again to
about 1.4 million. The same exercise can be repeated for other values
of the density quality, but also for the optical depth quality. The
bottom-line is always the same: for a given quality requirement, there
seems to be an optimal combination of $\delta_{\text{max}}$ and
$\tau_{\text{max}}$ that yields grids with a minimum number of
cells. Typically, the number of cells in this ideal combination is
reduced by 20\% compared to the grid based on a mass criterion only
(which corresponds to the asymptotic values of the contours towards the
right in the panels of Fig.~{\ref{Contours.fig}}).

These figures also demonstrate the superiority of the $k$-d
tree compared to the octree, also in the case when more advanced tree
construction criteria are adopted. Taking the same example as used
above, we see that the smallest octree with a density quality
$\Delta\rho=375$ contains around 1 million cells, whereas a $k$-d tree
with 800,000 cells can be constructed with the same quality.

\begin{figure*}
\centering
\includegraphics[width=0.45\textwidth]{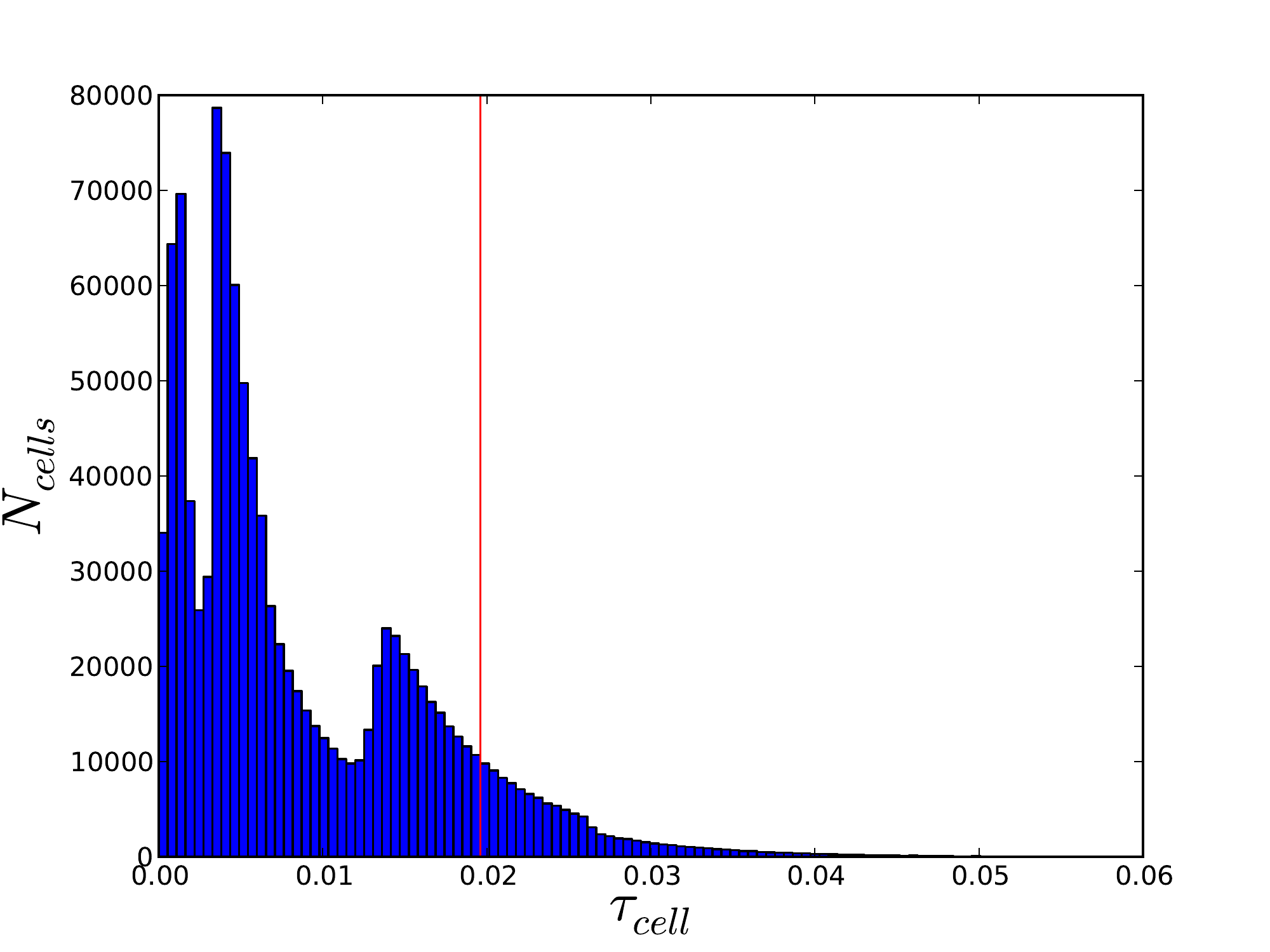}
\includegraphics[width=0.45\textwidth]{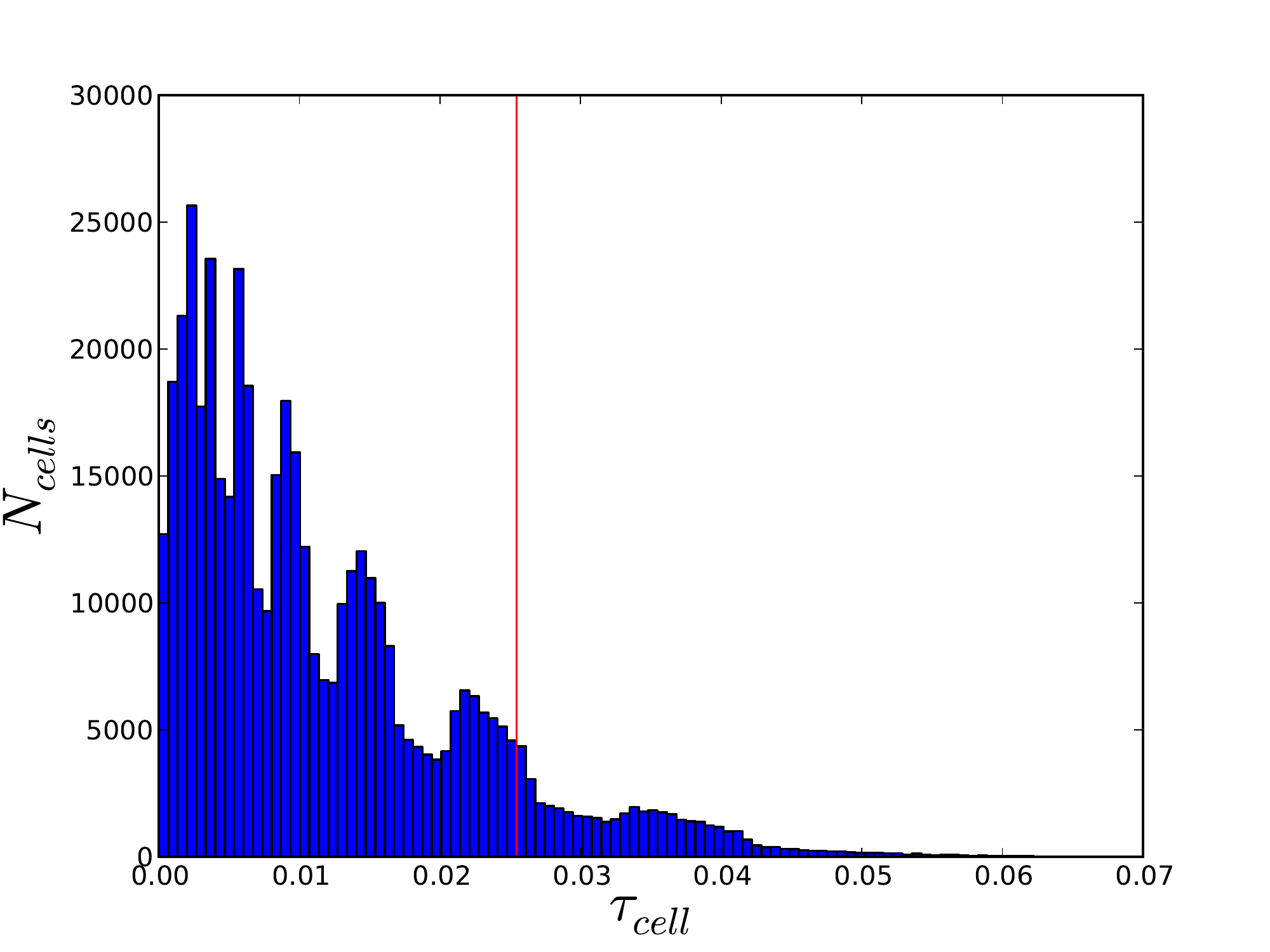}
\caption{Histograms of the distribution of the optical depth
  $\tau_{\text{cell}}$ in each cell for the spiral galaxy model octree
  grid (left panel) and $k$-d tree (right panel), corresponding to a
  mass threshold criterion $\delta_{\text{max}} = 3 \times 10^{-6}$. }
\label{OptDepthHisto.fig}
\end{figure*}

The question is now how the two criteria need to be combined to give
the optimal results. In other words, how do we set the values
$\delta_{\text{max}}$ and $\tau_{\text{max}}$ such that this
combination is ideal~? In our parameter space study, we can identify
these point by searching along every iso-quality contour the point
that corresponds to the smallest number of grid cells. These points
are indicated in red in the different panels of
Fig.~{\ref{Contours.fig}}. However, it is obviously impossible to
construct this entire parameter space of grids for every possible
radiative transfer simulation. Instead, it would be useful to have a
simple recipe that can identify this combination. A simple numerical
relation between $\delta_{\text{max}}$ and $\tau_{\text{max}}$ seems
impossible, as the useful range of $\tau_{\text{max}}$ values depends
on the total mass, the geometric complexity, and the opacity of our
model.

Our approach consists of  two steps. As a first step, we fix
the value of $\delta_{\text{max}}$ and construct a grid using this
subdivision stopping criterion only. Subsequently, we look at the
distribution of the optical depth of the
cells. Fig.~{\ref{OptDepthHisto.fig}} shows the histograms of cell
optical depth for the logarithmic spiral galaxy models for a fixed
$\delta_{\text{max}}=3\times10^{-6}$, for the octree and $k$-d
tree grids respectively. In both histograms, there is a long tail of
high optical depth, but not necessarily high mass, cells which are the
prime candidates for further subdivision. The vertical lines in these
plots indicates the 90\% percent of the optical depth distribution,
which we found to be a suitable value as to where this distribution
should be cut off. In order words, we create an ordered list of the
$\tau$ values, and we set $\tau_{\text{max}}$ equal to the optical
depth of the cell at position 0.9\,$N_{\text{cells}}$. The combinations
of $\delta_{\text{max}}$ and $\tau_{\text{max}}$ we have obtained in
this way are indicated as green dots in the different panels of
Fig.~{\ref{Contours.fig}}, and they lie very close to the optimal
points recovered from searching the parameter space.

This recipe of finding the optimal value of $\tau_{\text{max}}$
corresponding to a given value of $\delta_{\text{max}}$ is simple and
fast, as the necessary calculations are easily done during the tree
construction phase. The recipe is exactly the same for both octree and
$k$-d tree based grids. Our conclusion is that an additional optical
depth criterion, chosen according to a simple recipe, is very useful
in producing octree or $k$-d trees with fewer cells for a given
quality requirement, or vice versa, with higher quality for a fixed
number of grid cells.

\subsection{Strong gradients and sharp edges}

\begin{figure*} 
\centering
\includegraphics[width=0.33\textwidth]{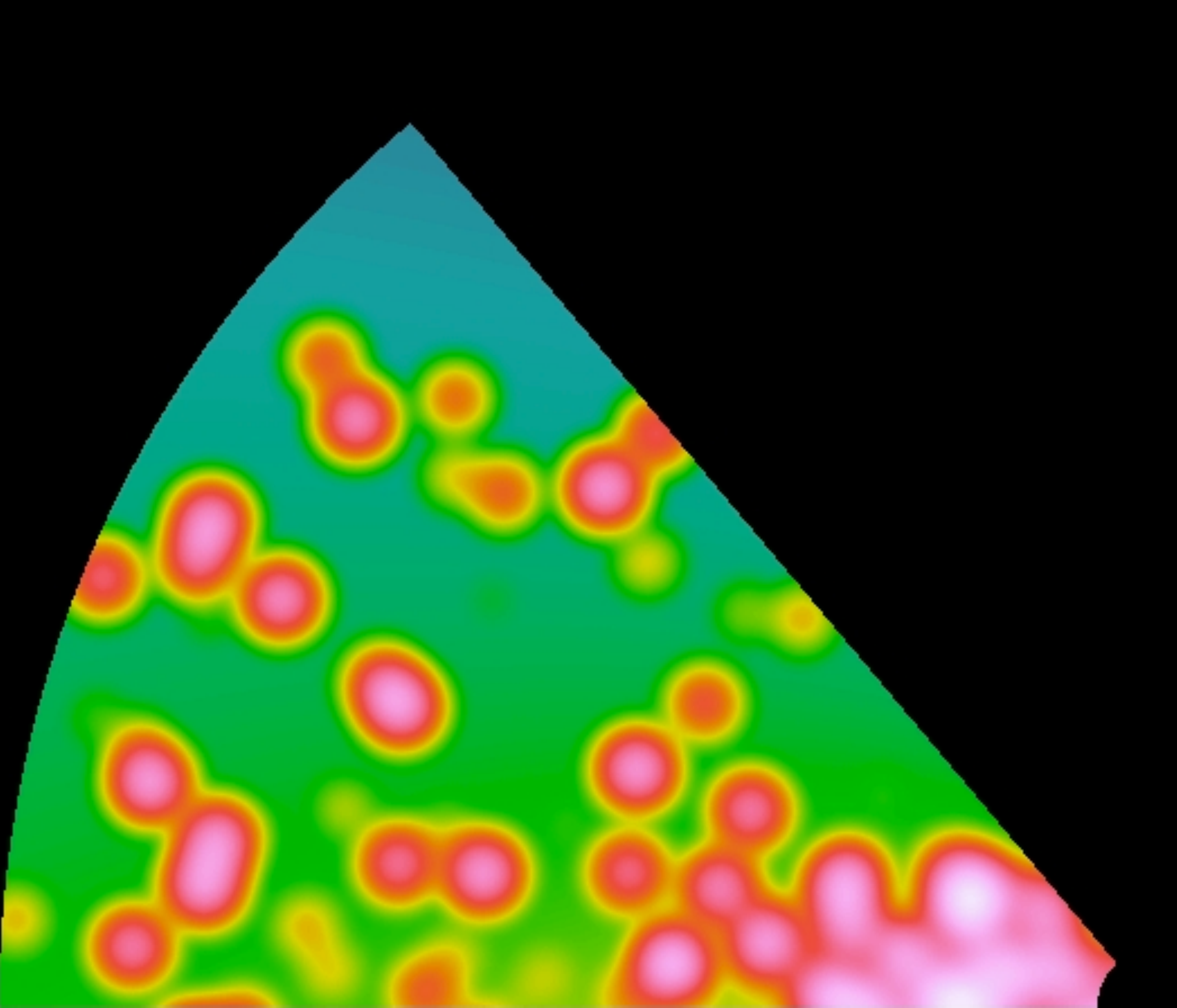} 
\includegraphics[width=0.33\textwidth]{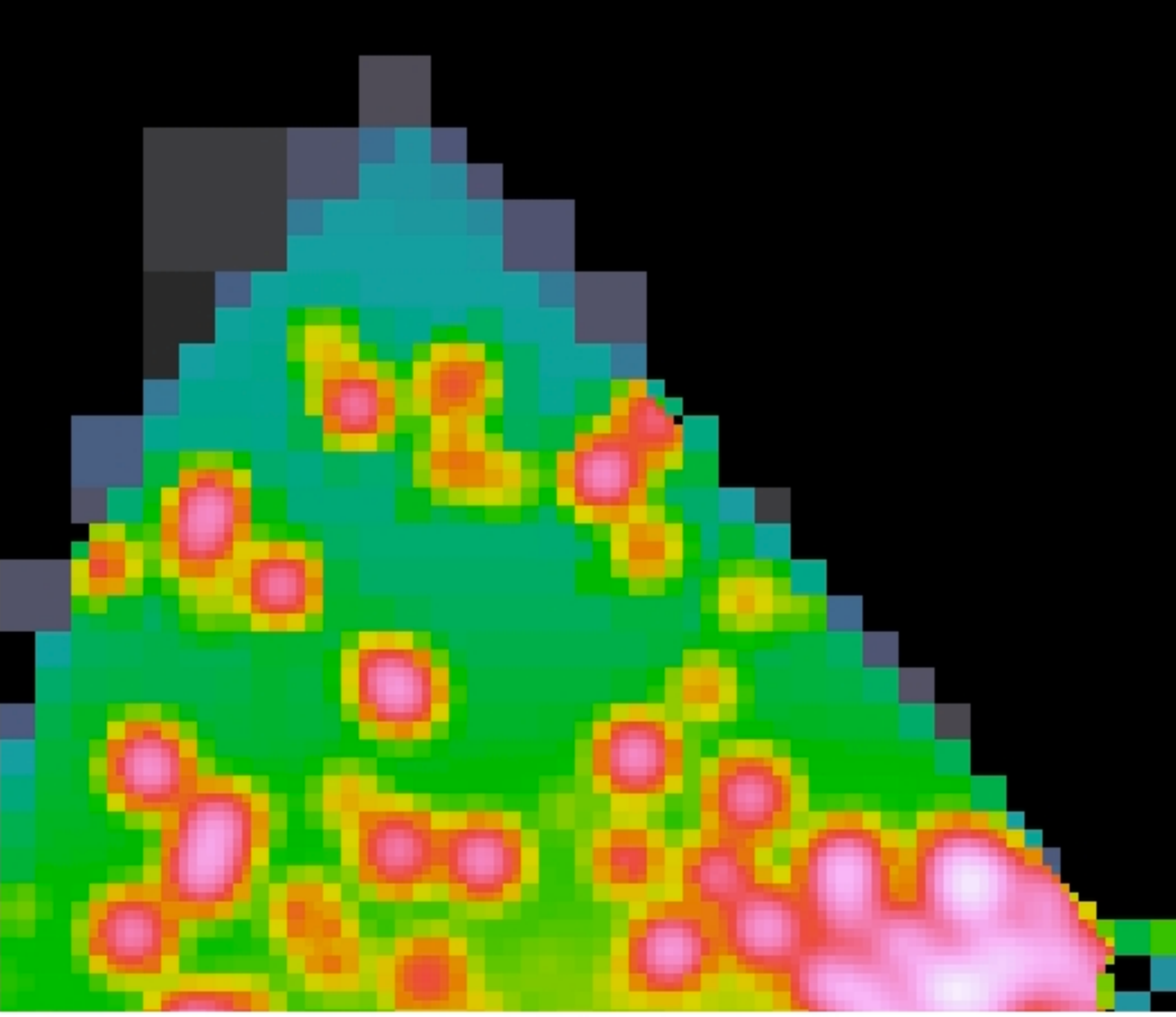} 
\includegraphics[width=0.33\textwidth]{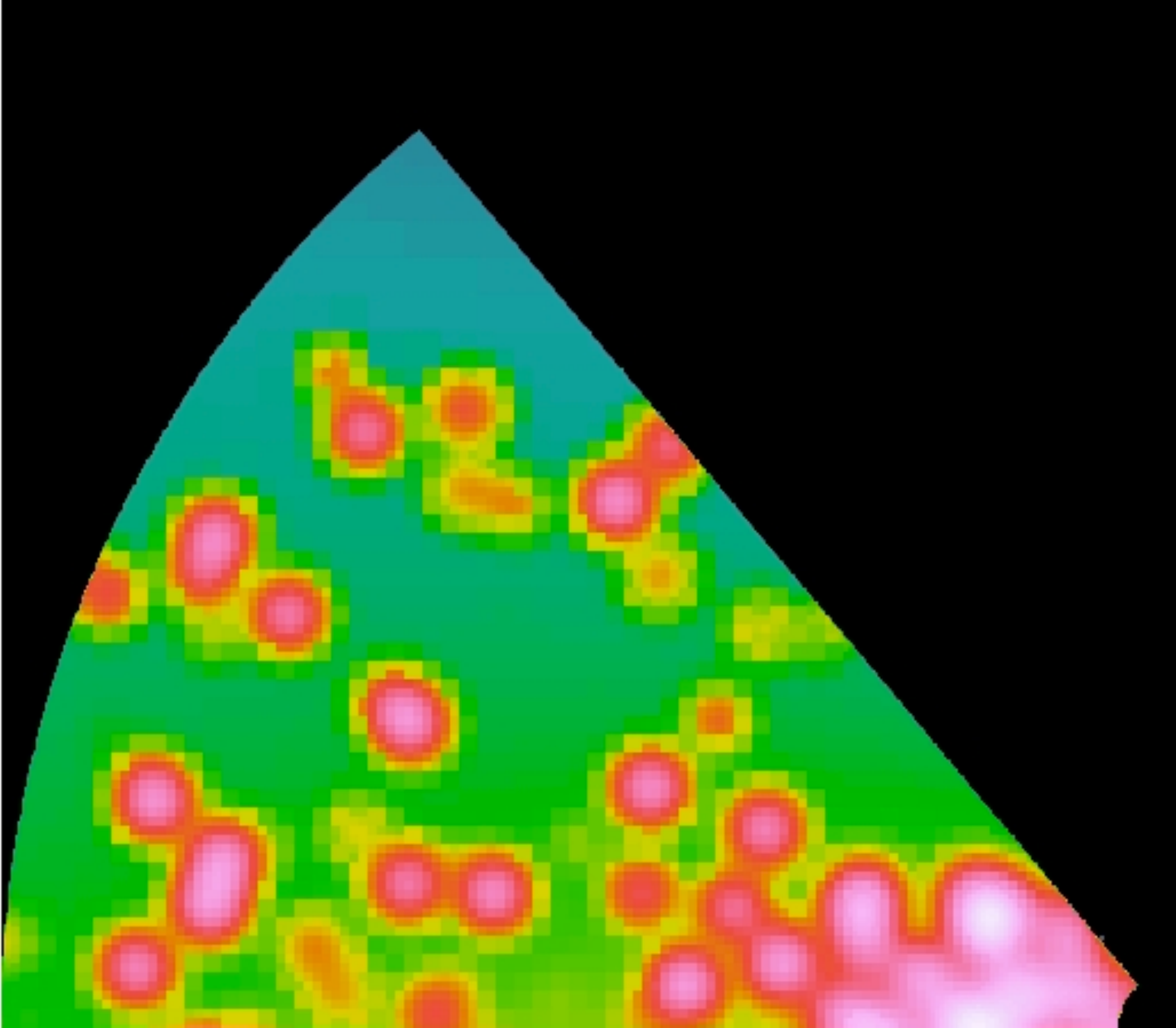} 
\caption{Illustration on how the density dispersion criterion
  $q_{\text{max}}$ solves the problem of sharp edges in models such as
  the AGN torus. In this example we used an octree grid. The left
  panel represents a cut through the true dust density in $xz$ plane,
  and the central panel shows the grid density as obtained without a
  density dispersion criterion with about 3 million
  cells. The edges of the torus are clearly poorly resolved. The
  right panel shows the grid density after adding the density
  dispersion criterion $q_{\text{max}} = 0.99$ and a
  maximum subdivision level of 10. The edges are
  now extremely well resolved using about 7 million of
  additional cells.}
\label{Torus.fig}
\end{figure*}

The other problem we referred to in the Introduction is the issue of
very strong gradients or sharp boundaries in the density fields. Grids
based on a mass criterion (or the combination of a mass and optical
depth criterion) have difficulties in dealing with these. A clear
example is the octree based grid for the clumpy AGN model considered
by \citet{2013A&A...554A..10S}. This model is characterized by a
density field with sharp boundaries at the edges of the torus. During
the tree construction process, we encounter many nodes at the edges of
this boundary, that only have a tiny and compact corner filled with
dust. When we compute the dust mass in such a node, it will soon be
below the dust mass (and/or optical depth) threshold, such that the
subdivision is stopped. This result is relatively large cells with a
relatively low density, whereas the true underlying density field is
relatively high in one tiny corner and zero in most of the cell, see
the central panel in Figure~{\ref{Torus.fig}} . This automatic stopping
of subdivision is a problem, as the regions with strong density
gradients and/or sharp boundaries are exactly regions that we want to
resolve at high resolution.

One way to solve this problem is to introduce artificial boundaries in
the grid. In the case of the clumpy AGN model, one could construct a
hierarchical grid within the sharp boundaries of the torus
itself. This solution might work efficiently for a number of
analytical models, but it is not the general solution desired for
arbitrary applications, where the occurrence and location of sharp
boundaries and strong gradients is not always known a priori.

Our proposed solution is to introduce a node subdivision criterion
that depends on the density gradient within a node, and add this
criterion to the already applied dust mass (and optical depth)
threshold criteria. We have considered a very simple approach that
does not introduce a strong computational overhead. During the
construction of the grid, the mass density is evaluated in
$N_{\text{ran}}$ random positions in each node as part of the estimate
of the mass in that node. If a certain node is not to be subdivided
further according to the $\delta_{\text{max}}$ and/or
$\tau_{\text{max}}$ criteria, we compute the quantity
\begin{equation}
	q
	=
	\begin{cases}
	\;\dfrac{\rho_{\text{max}}-\rho_{\text{min}}}{\rho_{\text{max}}}
	&
	\quad\text{if $\rho_{\text{max}}>0$,}
	\\
	\;0
	&
	\quad\text{if $\rho_{\text{max}}=0$.}
	\end{cases}
\end{equation}
where $\rho_{\text{min}}$ and $\rho_{\text{max}}$ are the smallest and
largest sampled density values from the list of $N_{\text{ran}}$
sampled positions in the node. This quantity is a simple measure for
the uniformity of the density within the node: for a constant density
in the node, $q=0$, whereas $q$ approaches 1 if a
steep gradient is present. The additional criterion we impose is that
a node is subdivided if $q$ exceeds a preset maximum value
$q_{\text{max}}\lesssim1$. In principle, nodes that contain a sharp
edge will continue to be subdivided for ever, since such cells have by
definition $q=1$. Thus it is important to always specify a reasonable
maximum subdivision level when using this subdivision criterion.

\begin{figure*}
\centering
\includegraphics[width=0.33\textwidth]{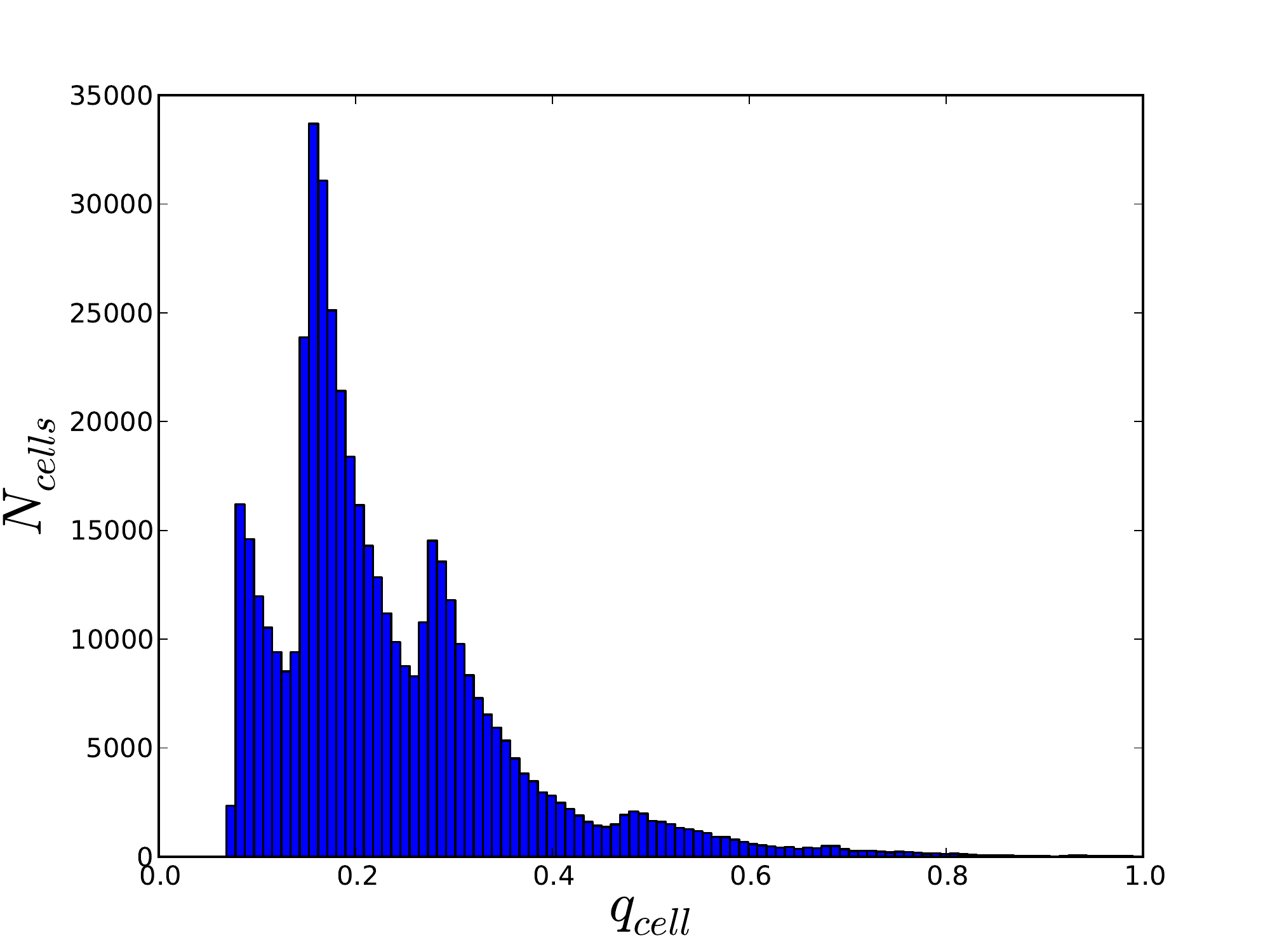}
\includegraphics[width=0.33\textwidth]{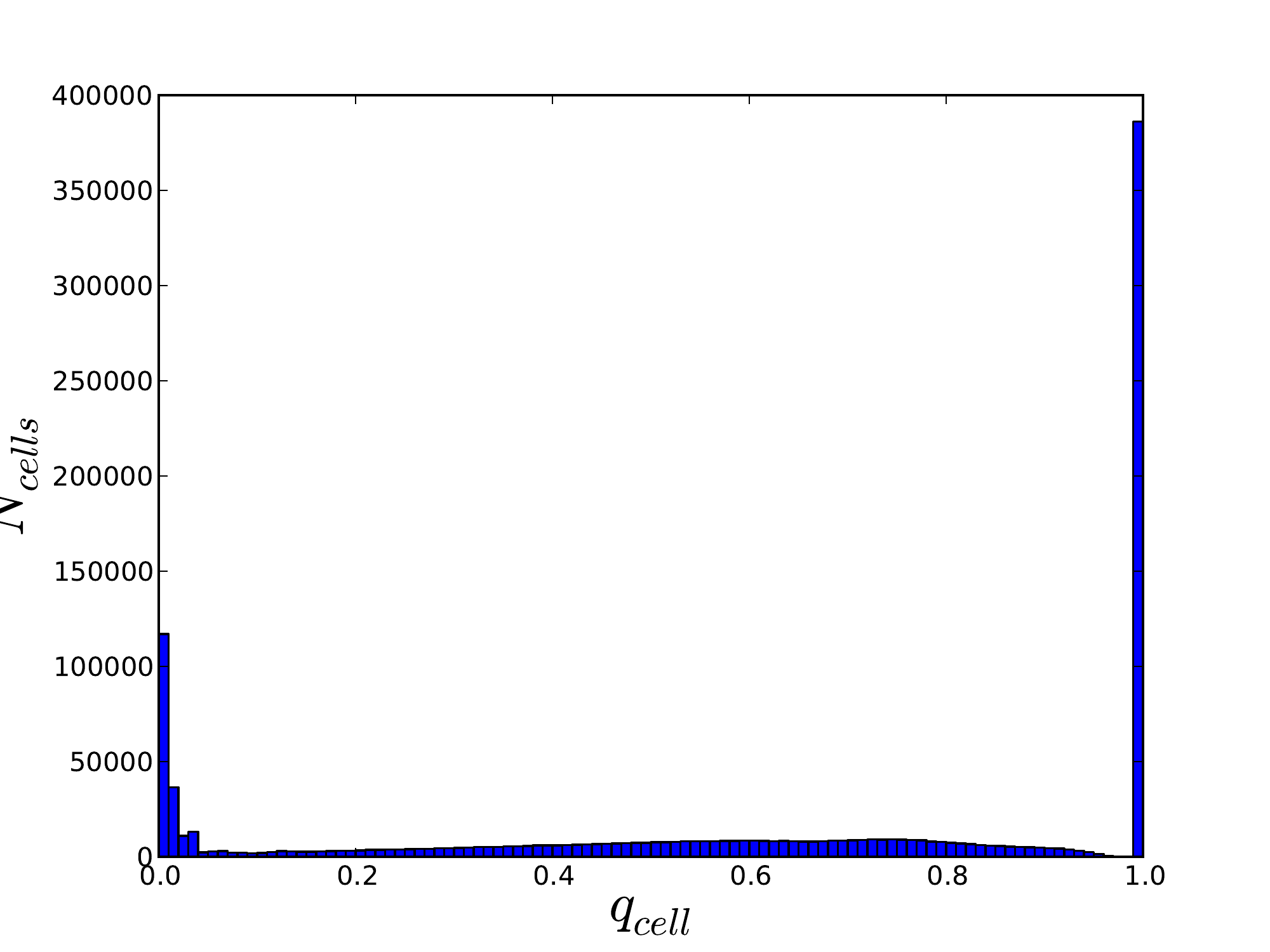}
\includegraphics[width=0.33\textwidth]{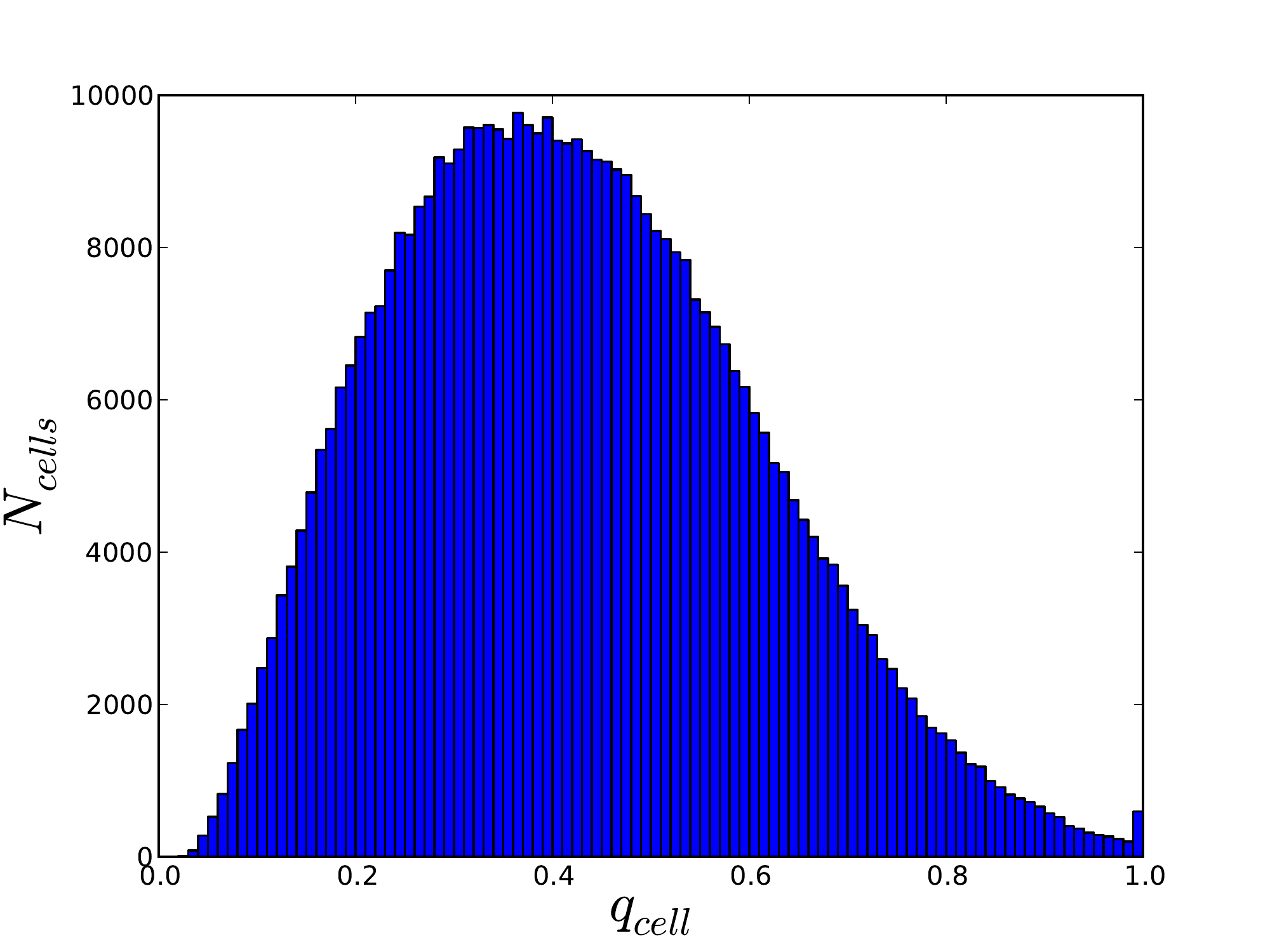}
\caption{Histograms of the distribution of the density dispersion $q$
  in each cell in our three models using the $k$-d tree grids,
  corresponding to a mass threshold criterion $\delta_{\text{max}} = 3
  \times 10^{-6}$. The left panel is the three-armed spiral galaxy
  model, the middle panel the clumpy AGN model, and the right
  panel the SPH galaxy model.}
  \label{DenDispHisto.fig}
\end{figure*}

In Fig.~{\ref{DenDispHisto.fig}} we show histograms of $q$ for $k$-d
tree based grids for the three test models we consider, before
applying any $q$-based subdivision. The logarithmic three-armed spiral
galaxy model (left panel) is the smoothest and most regular of the
three models, and is characterized by cells in which the vast majority
has $q<0.35$. The SPH galaxy model (right panel) is characterized by
a somewhat more irregular, but still relatively smooth density
field. The corresponding distribution of $q$ values is a fairly broad
distribution that peaks around 0.35 and then decreases smoothly
towards larger values of $q$. A small number of cells have $q=1$:
these correspond to cells at the edges of cavities in the dust
density. Finally, the clumpy AGN model (central panel) is designed to
be a test model with strong density gradients and sharp edges. This is
clearly evident in the histogram: a large fraction of the cells is
characterized by large values of $q$ and particularly conspicuous is
the large number of cells with $q=1$, corresponding to grid cells
overlapping with the sharp edge of the torus. In such a model, we
clearly need higher resolution for those
cells. Figure~{\ref{Torus.fig}} illustrates the differences in the
octree based grid for this model before and after adding the $q$-based
subdivision criterion. This simple recipe clearly solves the problem.

\section{Discussion and conclusion}
\label{Conclusion.sec}

The main goal of this paper was to critically investigate the
efficiency and accuracy of the standard octree algorithm in the context
of radiative transfer simulations, and to investigate the
use of alternative hierarchical grid structures and node subdivision
stopping criteria beyond a straightforward mass threshold value.

We have investigated the use of hierarchical $k$-d tree grids as an
alternative to octree grids to partition the transfer medium. We have
implemented a flexible $k$-d tree structure in the 3D Monte Carlo code
SKIRT \citep{2003MNRAS.343.1081B, 2011ApJS..196...22B}. The
construction algorithm of the $k$-d tree is completely similar to the
technique we used for the construction of the octree grid in SKIRT,
and we can use the same, very efficient, neighbor list search grid
traversal method. Using three different test models, and a set of
objective grid quality metrics, we have critically compared the octree
and $k$-d based grids. We have found that, for a fixed value of the
mass threshold $\delta_{\text{max}}$, the $k$-d tree generates only
half as many cells as the octree, irrespective of the chosen
model. Moreover, if we compare the quality of the octree and $k$-d
tree grids for a fixed total number of cells, the latter have a higher
accuracy. We can generate a $k$-d tree with roughly 20\% fewer cells
compared to an octree grid for a certain required density or optical
depth quality.

As a second objective, we have investigated whether there are useful
alternatives to the simple mass threshold as a criterion to stop the
recursive subdivision of the nodes in hierarchical trees (both octrees
and $k$-d trees). In order to stimulate the subdivision of small,
high-density cells, we have tested the option of an optical depth
criterion, but this gave rise to grids with an unacceptably low
accuracy. A combination of a mass and an optical depth criterion,
however, turned out to be a sensible criterion. Using an optimal
combination of both, it was possible to reduce the number of cells in
the grids by 20\% for a given quality requirement, and also decrease
the average number of cells crossed on a path, resulting in faster run
time of the simulations. We have presented a simple recipe that
enables us to approximate this optimal combination without the need to
scan the entire parameter space.

Finally, we have considered the problem caused by discontinuities,
strong gradients or sharp edges. In hierarchical grids (octrees or
$k$-d trees) governed by mass and/or optical depth threshold criteria,
such features can lead to undesirable large, low-density cells,
whereas a strong subdivision is particularly needed in these cases. We
have presented a simple solution for this problem, in the form of an
additional gradient threshold criterion. The criterion is very simple
to implement and is shown to give the desired results.

The main results of this investigation are that, in the context of
radiative transfer simulations, we can strongly advocate
the use of the lesser-known $k$-d trees as an alternative to the
popular octrees, and that the combination of different subdivision
stopping criteria, rather than a simple mass criterion, can lead to
more efficient grids. While this is another step forward towards the
construction of the ideal grid, we are well aware that this is not the
end stage. A number of side notes are required.

The first caveat refers to the different grid quality metrics that we
have used in this paper. We have attempted to quantify the quality of
a grid using different criteria: the total number of cells, the
average number of cells crossed per path, and an estimate of the
standard deviation of the density and optical depth discretization
error. These quantities are an attempt to translate what one would
consider intuitively as a good grid to a quantitative measure by which
different grids can be compared. They are only based on the dust
density field, and they are relatively straightforward to
calculate. But the question remains whether these metrics are
sufficient to really measure whether one grid is better than the other
for radiative transfer simulations. The ideal spatial grid in
radiative transfer simulations should discretize the intensity of the
radiation field as accurately as possible, but this quantity is
unknown at the start of the simulation (it is exactly the goal of
radiative transfer simulation to recover the radiation field). As the
radiation field is not only determined by the dust density, but
equally well by the distribution of sources and the optical properties
of the dust, it is simply impossible to construct the ideal grid with
only knowledge of the dust density field, or to construct grid quality
metrics based on the density field alone that measure the
appropriateness of a grid. As suggested by
\citet{2013A&A...554A..10S}, a more advanced option could in principle
be achieved iteratively from running a series of radiative transfer
simulations, in which the grid is adapted at every step based on the
properties of the radiation field (strength of the radiation field,
temperature distribution,\ldots) in the previous iteration step
\citep[see also][]{2006A&A...456....1N}. This approach is clearly too
complex and time-consuming for every single simulation. Grid
construction guidelines and quality metrics based on the dust density
field alone, such as discussed in this paper, are probably a good
compromise between effectiveness and computational cost, but future
work might investigate this in more detail.

The second side note is that the grids we have considered here
constitute only a small corner of the entire zoo of possible grid
structures that could be considered. Octrees and $k$-d trees are only
two members of the large family of hierarchical grids that have been
developed as space partitioning structures. In the field of
computational geometry and computer graphics, a wider range of space
partitioning structures are used for different goals, and hierarchical
AMR grids (or the so-called grids-in-grids) are also used in
several radiative transfer codes \citep[e.g.][]{2003CoPhC.150...99W,
  2006ApJ...645..792T, 2009ApJ...696..853L, 2011A&A...536A..79R,
  2012ApJ...751...27H, 2012A&A...544A..52L}. \citet{Havran2000} did an
extensive comparison of heuristic ray-shooting algorithms based on
many different space partitioning structures (BSP trees, $k$-d trees,
octrees, bounding volume hierarchies, uniform grids, and three types
of hierarchical grids) and found that the ray-shooting algorithm based
on the $k$-d tree is the winning candidate among all tested
algorithms. This gives us some confidence that, among the class of
hierarchical grids, the $k$-d tree is an ideal candidate for 
radiative transfer grids, but other candidates might
have their own advantages.

The same might also be true for unstructured grids, which belong to a
different section of the family of space partitioning
structures. Particularly interesting are unstructured grids based on
Voronoi or Delaunay tessellation. These grids have existed for a long
time, but they have recently gained increases popularity thanks to the
development of hydrodynamics codes that operate on them
\citep{1997MNRAS.288..903X, 2010MNRAS.401..791S,
  2011ApJS..197...15D}. Radiative transfer on such grids has been
shown to be possible \citep{2010A&A...515A..79P, 2010A&A...523A..25B,
  Camps2013}. While unstructured grids can probably not compete with
hierarchical grids in terms of grid traversal, they could be more
efficient in some situations, e.g.\ in simulations where in-cell
operations (such as the calculation of the temperature distribution or
the dust emission profile), rather than grid traversal, are the most
expensive operation.

\begin{acknowledgement}
  WS acknowledges the support of Al-Baath University and The Ministry
  of High Education in Syria in the form of a research grant. This
  work fits in the CHARM framework (Contemporary physical challenges
  in Heliospheric and AstRophysical Models), a phase VII
  Interuniversity Attraction Pole (IAP) programme organised by BELSPO,
  the BELgian federal Science Policy Office.
\end{acknowledgement}

\bibliographystyle{aa} 
\bibliography{KDtree}

\begin{thebibliography}{41}
\expandafter\ifx\csname natexlab\endcsname\relax\def\natexlab#1{#1}\fi

\bibitem[{{Baes} {et~al.}(2003){Baes}, {Davies}, {Dejonghe}, {Sabatini},
  {Roberts}, {Evans}, {Linder}, {Smith}, \& {de Blok}}]{2003MNRAS.343.1081B}
{Baes}, M., {Davies}, J.~I., {Dejonghe}, H., {et~al.} 2003, \mnras, 343, 1081

\bibitem[{{Baes} {et~al.}(2011){Baes}, {Verstappen}, {De Looze}, {Fritz},
  {Saftly}, {Vidal P{\'e}rez}, {Stalevski}, \& {Valcke}}]{2011ApJS..196...22B}
{Baes}, M., {Verstappen}, J., {De Looze}, I., {et~al.} 2011, \apjs, 196, 22

\bibitem[{Bentley(1975)}]{Bentley1975}
Bentley, J.~L. 1975, Commun. ACM, 18, 509

\bibitem[{{Bianchi}(2008)}]{2008A&A...490..461B}
{Bianchi}, S. 2008, \aap, 490, 461

\bibitem[{{Brinch} \& {Hogerheijde}(2010)}]{2010A&A...523A..25B}
{Brinch}, C. \& {Hogerheijde}, M.~R. 2010, \aap, 523, A25

\bibitem[{{Camps} {et~al.}(2013){Camps}, {Baes}, \& {Saftly}}]{Camps2013}
{Camps}, P., {Baes}, M., \& {Saftly}, W. 2013, \aap, submitted

\bibitem[{{Duffell} \& {MacFadyen}(2011)}]{2011ApJS..197...15D}
{Duffell}, P.~C. \& {MacFadyen}, A.~I. 2011, \apjs, 197, 15

\bibitem[{Fuchs {et~al.}(1979)Fuchs, Kedem, \& Naylor}]{Fuchs1979}
Fuchs, H., Kedem, Z.~M., \& Naylor, B. 1979, SIGGRAPH Comput.\ Graph., 13, 175

\bibitem[{Fuchs {et~al.}(1980)Fuchs, Kedem, \& Naylor}]{Fuchs1980}
Fuchs, H., Kedem, Z.~M., \& Naylor, B.~F. 1980, SIGGRAPH Comput.\ Graph., 14,
  124

\bibitem[{{Glassner}(1984)}]{Glassner1984}
{Glassner}, A.~S. 1984, IEEE Comput.\ Graph.\ Appl., 4, 15

\bibitem[{{Gordon} {et~al.}(2001){Gordon}, {Misselt}, {Witt}, \&
  {Clayton}}]{2001ApJ...551..269G}
{Gordon}, K.~D., {Misselt}, K.~A., {Witt}, A.~N., \& {Clayton}, G.~C. 2001,
  \apj, 551, 269

\bibitem[{{Harries} {et~al.}(2004){Harries}, {Monnier}, {Symington}, \&
  {Kurosawa}}]{2004MNRAS.350..565H}
{Harries}, T.~J., {Monnier}, J.~D., {Symington}, N.~H., \& {Kurosawa}, R. 2004,
  \mnras, 350, 565

\bibitem[{Havran(2000)}]{Havran2000}
Havran, V. 2000, Phd thesis, Czech Technical University

\bibitem[{{Heymann} \& {Siebenmorgen}(2012)}]{2012ApJ...751...27H}
{Heymann}, F. \& {Siebenmorgen}, R. 2012, \apj, 751, 27

\bibitem[{{Jonsson}(2006)}]{2006MNRAS.372....2J}
{Jonsson}, P. 2006, \mnras, 372, 2

\bibitem[{{Kurosawa} \& {Hillier}(2001)}]{2001A&A...379..336K}
{Kurosawa}, R. \& {Hillier}, D.~J. 2001, \aap, 379, 336

\bibitem[{{Laursen} {et~al.}(2009){Laursen}, {Razoumov}, \&
  {Sommer-Larsen}}]{2009ApJ...696..853L}
{Laursen}, P., {Razoumov}, A.~O., \& {Sommer-Larsen}, J. 2009, \apj, 696, 853

\bibitem[{{Lunttila} \& {Juvela}(2012)}]{2012A&A...544A..52L}
{Lunttila}, T. \& {Juvela}, M. 2012, \aap, 544, A52

\bibitem[{MacDonald \& Booth(1990)}]{MacDonald1990}
MacDonald, D.~J. \& Booth, K.~S. 1990, Vis.\ Comput., 6, 153

\bibitem[{{Misiriotis} {et~al.}(2000){Misiriotis}, {Kylafis}, {Papamastorakis},
  \& {Xilouris}}]{2000A&A...353..117M}
{Misiriotis}, A., {Kylafis}, N.~D., {Papamastorakis}, J., \& {Xilouris}, E.~M.
  2000, \aap, 353, 117

\bibitem[{{Niccolini} \& {Alcolea}(2006)}]{2006A&A...456....1N}
{Niccolini}, G. \& {Alcolea}, J. 2006, \aap, 456, 1

\bibitem[{{Paardekooper} {et~al.}(2010){Paardekooper}, {Kruip}, \&
  {Icke}}]{2010A&A...515A..79P}
{Paardekooper}, J.-P., {Kruip}, C.~J.~H., \& {Icke}, V. 2010, \aap, 515, A79

\bibitem[{{Pinte} {et~al.}(2006){Pinte}, {M{\'e}nard}, {Duch{\^e}ne}, \&
  {Bastien}}]{2006A&A...459..797P}
{Pinte}, C., {M{\'e}nard}, F., {Duch{\^e}ne}, G., \& {Bastien}, P. 2006, \aap,
  459, 797

\bibitem[{{Rahimi} \& {Kawata}(2012)}]{2012MNRAS.422.2609R}
{Rahimi}, A. \& {Kawata}, D. 2012, \mnras, 422, 2609

\bibitem[{Reshetov {et~al.}(2005)Reshetov, Soupikov, \& Hurley}]{Reshetov2005}
Reshetov, A., Soupikov, A., \& Hurley, J. 2005, ACM Trans.\ Graph., 24, 1176

\bibitem[{{Robitaille}(2011)}]{2011A&A...536A..79R}
{Robitaille}, T.~P. 2011, \aap, 536, A79

\bibitem[{{Saftly} {et~al.}(2013){Saftly}, {Camps}, {Baes}, {Gordon},
  {Vandewoude}, {Rahimi}, \& {Stalevski}}]{2013A&A...554A..10S}
{Saftly}, W., {Camps}, P., {Baes}, M., {et~al.} 2013, \aap, 554, A10

\bibitem[{{Schechtman-Rook} {et~al.}(2012){Schechtman-Rook}, {Bershady}, \&
  {Wood}}]{2012ApJ...746...70S}
{Schechtman-Rook}, A., {Bershady}, M.~A., \& {Wood}, K. 2012, \apj, 746, 70

\bibitem[{Shevtsov {et~al.}(2007)Shevtsov, Soupikov, \&
  Kapustin}]{Shevtsov2007}
Shevtsov, M., Soupikov, A., \& Kapustin, A. 2007, Comput. Graph. Forum, 26, 395

\bibitem[{{Springel}(2010)}]{2010MNRAS.401..791S}
{Springel}, V. 2010, \mnras, 401, 791

\bibitem[{{Stalevski} {et~al.}(2012){Stalevski}, {Fritz}, {Baes}, {Nakos}, \&
  {Popovi{\'c}}}]{2012MNRAS.420.2756S}
{Stalevski}, M., {Fritz}, J., {Baes}, M., {Nakos}, T., \& {Popovi{\'c}}, L.~{\v
  C}. 2012, \mnras, 420, 2756

\bibitem[{{Steinacker} {et~al.}(2006){Steinacker}, {Bacmann}, \&
  {Henning}}]{2006ApJ...645..920S}
{Steinacker}, J., {Bacmann}, A., \& {Henning}, T. 2006, \apj, 645, 920

\bibitem[{{Steinacker} {et~al.}(2013){Steinacker}, {Baes}, \&
  {Gordon}}]{2013ARA&A..51...63S}
{Steinacker}, J., {Baes}, M., \& {Gordon}, K.~D. 2013, \araa, 51, 63

\bibitem[{{Tasitsiomi}(2006)}]{2006ApJ...645..792T}
{Tasitsiomi}, A. 2006, \apj, 645, 792

\bibitem[{Wald \& Havran(2006)}]{Wald2006}
Wald, I. \& Havran, V. 2006, in IEEE Symposium on Interactive Ray Tracing,
  61--70

\bibitem[{{Whitney}(2011)}]{2011BASI...39..101W}
{Whitney}, B.~A. 2011, Bulletin of the Astronomical Society of India, 39, 101

\bibitem[{{Whitney} {et~al.}(2013){Whitney}, {Robitaille}, {Bjorkman}, {Dong},
  {Wolff}, {Wood}, \& {Honor}}]{2013ApJS..207...30W}
{Whitney}, B.~A., {Robitaille}, T.~P., {Bjorkman}, J.~E., {et~al.} 2013, \apjs,
  207, 30

\bibitem[{{Wolf}(2003)}]{2003CoPhC.150...99W}
{Wolf}, S. 2003, Computer Physics Communications, 150, 99

\bibitem[{{Xilouris} {et~al.}(1997){Xilouris}, {Kylafis}, {Papamastorakis},
  {Paleologou}, \& {Haerendel}}]{1997A&A...325..135X}
{Xilouris}, E.~M., {Kylafis}, N.~D., {Papamastorakis}, J., {Paleologou}, E.~V.,
  \& {Haerendel}, G. 1997, \aap, 325, 135

\bibitem[{{Xu}(1997)}]{1997MNRAS.288..903X}
{Xu}, G. 1997, \mnras, 288, 903

\bibitem[{Zhou {et~al.}(2008)Zhou, Hou, Wang, \& Guo}]{Zhou2008}
Zhou, K., Hou, Q., Wang, R., \& Guo, B. 2008, ACM Trans. Graph., 27, 126

\end{thebibliography}

\end{document}